\documentclass[useAMS,usenatbib]{mnras}
\usepackage{times}
\usepackage{color}
\usepackage{breakurl}
\usepackage{color}
\usepackage{mathrsfs}

\usepackage{multicol}

\usepackage{bm}		

\usepackage{pdflscape}	

\usepackage{amsfonts,amsmath,amssymb,mathrsfs}
\usepackage{graphicx,epsf,epsfig}

\usepackage{inputenc}
\usepackage{subfigure}
\usepackage{bm}
\usepackage{longtable}
\usepackage[usenames,dvipsnames]{xcolor}
\usepackage{multirow}

\usepackage{hyperref}
\usepackage{times}
\usepackage{color}

\usepackage{appendix}

\def\be{\begin{equation}}
\def\ee{\end{equation}}
\def\bea{\begin{eqnarray}}
\def\eea{\end{eqnarray}}

\title[Reconstruction of accelerated-decelerated phase]{Model independent reconstruction of cosmological accelerated-decelerated phase}

\author[S.~Capozziello, Peter~K.S.~Dunsby and O.~Luongo]{Salvatore Capozziello$^{1,2,3,4}$ Peter K.S. Dunsby,$^{5,6,7}$ and Orlando Luongo$^{8,9,10}$\thanks{orlando.luongo@unicam.it}  \\
$^1$Dipartimento di Fisica E. Pancini, Universit\`a di Napoli "Federico II", Via Cinthia, I-80126 Napoli, Italy.\\
$^2$Scuola Superiore Meridionale, Largo S. Marcellino 10, I-80138, Napoli, Italy.\\
$^3$Istituto Nazionale di Fisica Nucleare (INFN), Sezione di Napoli, Via Cinthia, Napoli, Italy.\\
$^4$Laboratory of Theoretical Cosmology, Tomsk State University of Control Systems and Radioelectronics (TUSUR), Tomsk, Russia.\\
$^5$ Department of Mathematics and Applied Mathematics, University of Cape Town, Rondebosch 7701, Cape Town, South Africa.\\
$^6$Astrophysics, Cosmology and Gravity Centre (ACGC),
University of Cape Town, Rondebosch 7701, Cape Town, South Africa.\\
$^7$South African Astronomical Observatory, Observatory 7925, Cape Town, South Africa.\\
$^8$ School of Science and Technology, University of Camerino, Via Madonna delle Carceri 9, 62032, Camerino, Italy.\\
$^9$Dipartimento di Matematica, Universit\`a di Pisa, Largo B. Pontecorvo 5, Pisa, 56127, Italy.\\
$^{10}$ Institute of Experimental and Theoretical Physics, Al-Farabi Kazakh National University, Almaty 050040, Kazakhstan.
}

\begin{document}

\maketitle

\begin{abstract}
We propose two model independent methods to obtain constraints on the transition and equivalence redshifts $z_{tr}$, $z_{eq}$. In particular, we consider $z_{tr}$ as the onset of cosmic acceleration, whereas $z_{eq}$ the redshift at which the densities of dark energy and pressureless matter are equated. With this prescription, we expand the Hubble and deceleration parameters up to two hierarchical orders and   show a linear correlation between transition and equivalence, from which we propose exclusion plots where $z_{eq}$ is not allowed to span. To this end, we discuss how to build up cosmographic expansions in terms of $z_{tr}$ and compute the corresponding observable quantities directly fitting the luminosity and angular distances and the Hubble rate with cosmic data. We make our computations through Monte Carlo fits involving type Ia supernova, baryonic acoustic oscillation and Hubble most recent data catalogs. We show at $1\sigma$ confidence level the $\Lambda$CDM predictions on $z_{tr}$ and $z_{eq}$ are slightly confirmed, although at $2\sigma$ confidence level dark energy expectations cannot be excluded. Finally, we theoretically interpret our outcomes and discuss possible limitations of our overall approach.
\end{abstract}



\section{Introduction}\label{S1:Introduction}

The standard cosmological model is plagued by a high degree of physical uncertainty due to the existence of dark energy and dark matter whose nature is currently the object of much theoretical debate \citep{review1,review2}. Arguably the greatest puzzle of the standard cosmological model is the prospect that $75\%$ of universe's content is made up of dark energy, responsible for the cosmic speed up \citep{luongo2}, whereas dark matter contributes for about the $20\%$ as an elusive component needed to guarantee structure formation\footnote{There is a growing need of new dark matter candidates as the most popular ones, e.g., weakly interacting massive particles, axions, and sterile neutrinos have not been detected so far. A possible different perspective on dark matter's origin is provided by \citep{luongo1}.}. A complete understanding of both these two constituents represents a real challenge of the the standard cosmological model, hereafter referred to as the $\Lambda$CDM paradigm \citep{review3}.

This approach involves treating dark energy as a vacuum energy cosmological constant, $\Lambda$, derived from quantum field fluctuations \citep{weinberg} and interpreted as a fluid that uniformly fills all of space. In the current epoch, the $\Lambda$CDM paradigm requires at least pressureless matter, negligible radiation and spatial curvature  \citep{review2}, with an overall negative pressure, determining the bizarre repulsive gravity \citep{Lambda} that accelerates the universe at the \emph{transition time}. The $\Lambda$CDM scenario is plagued by different issues that limit its theoretical interpretation. First, both $\Lambda$ and matter magnitudes are unexpectedly comparable, leading to a \textit{coincidence problem}. Second, constraints over $\Lambda$ turns out to be impressively smaller than the predicted vacuum energy density, providing a \textit{fine tuning problem}. In turn, these caveats have spurred the search for other physical explanations to describe the dynamics of the universe today, leading to a cascade of alternative approaches \citep{casc}. Recent tensions, in particular, seem to  shed light on the need of a more general framework that extends the $\Lambda$CDM paradigm \citep{cit2,Spallicci,Micol}.

Several extensions are possible\footnote{Among others, e.g., the addition of dynamical scalar fields with slowly varying potentials, models with interacting dark matter and dark energy components, and modifications to general relativity on large scales have been proposed \citep{revluongo}.}, although none of these models represent concrete alternatives to the standard concordance paradigm. In particular, one always identifies two relevant epochs associated with any dark energy models: $I)$ the time at which dark energy and matter have the same density, namely the equivalence time, $2)$ the time of dark energy domination over matter, {\it i.e.}, the onset of acceleration - the transition time. One can quantitatively identify the two epochs, {\it i.e.}, the first epoch starts at an equivalence redshift,  $z_{eq}$, at which the equality between matter, $\rho_m$, and dark energy, $\rho_X$, formally holds, namely $\rho_{m}(z_{eq})= \rho_X(z_{eq})$. The second epoch corresponds to the transition redshift, namely $z_{tr}$, at which the universe enters a phase of accelerated expansion \citep{melchiorri,libro}.

Recently it has been argued that such redshifts can be considered as \emph{kinematic quantities}, in a sense that constraining their values in a model-independent manner provides a way to describe the universe dynamics without postulating the model \emph{a priori} \citep{cosmography1}. Commonly, claims on the dependence of these two quantities on cosmological models have been made in the literature (see for example \citep{melchiorri} and references therein). To characterise them without postulating the underlying cosmological model can significantly help to discriminate between different dark energy scenarios, in analogy to the strategy of modern cosmography \citep{cosmography2,cosmography3,cosmography4,cosmography5,brah1}.

In this work, we propose two strategies to obtain the transition and equivalence redshifts using the model-independent treatment offered by cosmography. We adopt two procedures, the first makes use of a direct Hubble expansion, whereas the second expands the deceleration parameter. In both the two methods, we assume expansions around the transition redshift, namely $z\simeq z_{tr}$. The corresponding cosmological distances can be reformulated accordingly and this naturally implies to confront our expansions directly with data. We assume two different hierarchies, based on the order at which the expansions are performed. In particular, we first expand up to the jerk  and then up to the snap parameters at $z=z_{tr}$,  respectively. Afterwards, we take into account two kinds of fits based on the use of observational Hubble data (OHD) first and then Pantheon type Ia supernovae (SNe IA), baryonic acoustic oscillarions (BAOs) and OHD, for a total of eight fits performed through Markov Chain Monte Carlo (MCMC) simulations by means of a modified free available \texttt{Wolfram Mathematica} code \citep{codice}, developed with the Metropolis-Hastings algorithm. We discuss multiplicative and additive degeneracies over our coefficients, and we debate about possible underestimated error bars and on the overall $z_{tr}$ and $z_{eq}$ accuracy. The corresponding jerk and snap terms at the transition are also obtained. We therefore interpret the feasibility of our numerical results and introduce exclusions regions for $z_{eq}$. To do so, we   provide intervals of validity for the set $(z_{tr}, z_{eq})$ and discuss how these terms can discriminate dark energy evolution and its nature, confronting our outcomes with the standard cosmological model.

The paper is structured as follows. In Sect. \ref{sec:method}, we present the main features of the transition and equivalence redshifts in view of our cosmographic approach. In Sect. \ref{sez3}, we propose our two treatments adopted throughout this work in order to determine bounds on $z_{tr}$ and $z_{eq}$. In Sects. \ref{metodo} and \ref{sez4} we first propose the numerical methodology that we adopt for our fits and then we highlight our experimental procedures. We also propose our exclusion regions for $z_{eq}$. Once our results have been estimated, in Sect. \ref{sez6} we provide our theoretical interpretation, {\it i.e.}, we discuss the limitations and consequences of our recipe,  together with the theoretical consequences of our findings on dark energy. Finally, in Sect. \ref{sez7} we present our conclusions and perspectives. Details on cosmographic series, observational data and contour plots are given in Appendices A, B and C, respectively.

\section{Reconstructing cosmographic transition and equivalence redshifts}
\label{sec:method}

In this section, we describe the main features of the transition and equivalence redshifts, $z_{tr}$ and $z_{eq}$. We highlight the most important properties of such quantities and  fix them for the standard cosmological model, the $\Lambda$CDM paradigm, and for a generic class of dark energy models, where we do not \emph{a priori} postulate the underlying dark energy evolution.

To do so, we need to add extra conditions to ensure the  consistency of our  theory with the cosmological principle. For instance, to correctly embed within it, in a homogeneous and isotropic universe, we take the total pressure as sum of all sub-pressures of each constituents, {\it i.e.}, $P=\sum_iP_i$. Thus,  at the level of the background cosmology, we neglect spatial curvature and assume pressureless matter, $P_m=0$, immediately providing $P=P_X$, where $P_X$ is dark energy's pressure. In analogy, we take the total density to be\footnote{The  $\Lambda$CDM model depends on six parameters:  baryon and cold dark matter densities, the age of the universe, the scalar spectral index, the curvature fluctuation amplitude and reionization optical depth. Our choice dramatically reduces this set to  $\{H_0,\rho_m\}$. Any dark energy models turn out to be more complicated, depending on a few extra terms. } $\rho=\rho_m+\rho_X$.

In addition, one way to work out what sort of onset of cosmic acceleration and equivalence between dark energy and matter we expect, let us examine two more hypotheses. The first relies on the fact $z_{tr}$ and $z_{eq}$ \emph{are not} free but rather they depend upon dark energy's parameters \citep{ra1,ra2}. Consequently, it is natural to admit a connection between $z_{eq}$ and $z_{tr}$. The second is the requirement $z_{eq}\leq z_{tr}$, since dark energy requires time to equate  matter first and then to dominate over it \citep{ra3,ra4}.

Thus, we adopt the homogeneous and isotropic Friedmann-Robertson-Walker universe, where the dynamics is dictated by the Friedmann equations
\begin{equation}\label{fry}
H^2 \equiv \dfrac{\dot{a}}{a}={8\pi G\over3}\rho,\quad \dot H + H^2=-{4\pi G\over3}\left(\rho+3P\right),
\end{equation}
in which we made the aforementioned assumption of a spatially flat universe. In this respect, a generic spatially-flat dark energy scenario, with negligible radiation and neutrino contributions, is characterized by a generic evolution of dark energy. In this respect, let us consider a dark energy density, namely $\rho_{DE}=\mathcal G f(z)$, with $\mathcal G$ a constant with $f(z)$ an arbitrary-defined dark energy function. Then, to guarantee that $H(z=0)=H_0$ the constant, $\mathcal G$ turns out to be $\mathcal G=1-\Omega_{m,0}$, when  $f(z=0)\equiv f_0=1$. This choice represents the simplest approach since we are excluding any coupling term between dark matter and dark energy. The largest variety of dark energy models falls within this class of frameworks and well-behaves at different epochs of the universe evolution. In other words, $f(z)$ is a smooth function that fully determines how dark energy evolves with time, fulfilling the standard experimental evidence for which $f(\infty)\lesssim 1$ in order to guarantee small perturbations at primordial times. Thus, plugging this information in the first Friedmann equation, we provide the Hubble evolution by \citep{DEgenerico}
\begin{equation}\label{associato}
\left(\dfrac{H}{H_0}\right)^2=\Omega_{m,0}(1+z)^3+\left(1-\Omega_{m,0}\right)f(z)\,,
\end{equation}
where $\Omega_{m,0}\equiv{\rho_{m,0}\over\rho_{cr}}$, $\rho_{cr}\equiv\frac{3H_0^2}{8\pi G}$. Under this assumption, differentiating with respect to $z$, one has
\begin{equation}\label{sop}
2\frac{1+q}{1+z}\mathcal E^2=3\frac{\Omega_{m}(z)}{1+z}+\left(1-\Omega_{m,0}\right)f^\prime(z),
\end{equation}
with $\mathcal E^2\equiv\left(\frac{H}{H_0}\right)^2$ and $\Omega_m(z)\equiv\Omega_{m,0}(1+z)^3$.

\noindent In Eq. \eqref{sop}, we introduced the deceleration parameter by
\begin{equation}\label{decpar}
q\equiv-\dfrac{1}{H^2}\dfrac{\ddot{a}}{a}=-1+\dfrac{1+z}{H}\ \dfrac{dH}{dz}\,,
\end{equation}
where we used $a=(1+z)^{-1}$ with $a(0)=a_0=1$. Note the prime indicates the derivative with respect to $z$, while the dot refers to time derivative.

\subsection{Relating transition to equivalence redshifts}\label{IIA}

The transition redshift defines the onset of cosmic speed up. Since before current time there exists a matter-dominated epoch, in order to ensure  dark energy to accelerate the universe today, we require the deceleration parameter to change its sign. The situation is clear assuming the r.h.s. of Eq. \eqref{fry}, where $\ddot a= \dot H+H^2$ and, by virtue of Eq. \eqref{decpar}, we immediately see
$\dot H+H^2=-qH^2=\ddot a$, indicating the cases $\ddot a>0$ and $\ddot a<0$ respectively for our time and matter domination.

Consequently, the coarse-grained condition to let the transition occur is
\begin{equation}\label{semov}
q(z_{tr})=0\,,
\end{equation}
guaranteeing the acceleration changes sign throughout the recent universe history.

Since $q$ vanishes at $z_{tr}$, using the differential expression \eqref{sop}, we get $$
2\mathcal E^2=3\Omega_{m}(z)+\left(1-\Omega_{m,0}\right)(1+z)f^\prime(z),$$
and at the transition time, we have $z=z_{tr}$. Thus, $\mathcal E^2|_{z=z_{tr}}\equiv \mathcal E_{tr}$, $f^\prime(z)|_{z=z_{tr}}\equiv f^\prime(z_{tr})$ and $\Omega_{m}(z)|_{z=z_{tr}}\equiv\Omega_{m}(z_{tr})$. Consequently, isolating the term $\propto (1+z)$ we have the following formal solution for $z_{tr}$
\begin{equation}\label{generictransition}
z_{tr}=\frac{2\mathcal E_{tr}^2-3\Omega_{m,tr}}{(1-\Omega_{m,0})f^\prime_{tr}}-1,
\end{equation}
clearly valid in the case $f^\prime\neq0$. Here, $\mathcal E_{tr}\equiv\frac{H(z_{tr})}{H_0}$ and  $\Omega_{m,tr}\equiv\Omega_{m,0}(1+z_{tr})^3$, whereas for $f^\prime=0$ we reduce to the $\Lambda$CDM model, {\it i.e.}, $f(z)=1$ and we get
\begin{equation}
z_{tr}=\left(\dfrac{2\Omega_\Lambda}{\Omega_{m,0}}\right)^{{1\over3}}-1\,,\label{eq:z_tr LCDM}
\end{equation}
with $\mathcal E=\sqrt{\Omega_{m,0}(1+z)^3+\Omega_\Lambda}$ and $\Omega_\Lambda=1-\Omega_{m,0}$.

For the equivalence redshift, $z_{eq}$, following an analogous strategy one infers the formal solution
\begin{equation}\label{genericequivalence}
z_{eq}=-1+\Big[\frac{(1-\Omega_{m,0})f(z_{eq})}{\Omega_{m,0}}\Big]^{\frac{1}{3}}\,.
\end{equation}
Immediately, for the concordance paradigm we compute
\begin{equation}
z_{eq}=\left(\dfrac{\Omega_\Lambda}{\Omega_{m,0}}\right)^{{1\over3}}-1\,.
\label{eq:z_eq LCDM}
\end{equation}

Confronting Eqs. \eqref{generictransition} and \eqref{genericequivalence}, or more easily Eqs. \eqref{eq:z_tr LCDM} and \eqref{eq:z_eq LCDM}, we can combine $z_{tr}$ with $z_{eq}$ to formally yield $z_{tr}=z_{tr}(z_{eq})$ and  arguing \emph{de facto} that there exists a \emph{correlation} between transition and equivalence, as we speculated above. The $\Lambda$CDM case provides
\begin{equation}
z_{tr}=2^{\frac{1}{3}}\,(1+z_{eq})-1\,.
\label{eq:z_tr-z_eq LCDM}
\end{equation}
With arbitrary dark energy models, it is reasonable to get the same functional behaviour. So, approximating at first order, a linear dependence between $z_{eq}$ and $z_{tr}$ can be motivated in analogy to the standard paradigm to give the following ansatz
\begin{equation}
z_{tr}=\alpha+\beta z_{eq}.
\label{eq:ansatz}
\end{equation}
We assume Eq. \eqref{eq:ansatz} to be reasonable from now on. Moreover, later in the text we find its validity for generic dark energy models by using our strategies of numerical analysis.  For instance, looking at Eqs. \eqref{eq:z_tr-z_eq LCDM} and \eqref{eq:ansatz}, we notice $\alpha=2^{\frac{1}{3}}-1$ and $\beta=2^{\frac{1}{3}}$. The phase space of $\{\alpha,\beta\}$ is not arbitrary, since the condition $z_{tr}-z_{eq}\neq0$ holds. So, it appears evident that not only $\beta>0$ but also $\beta>1$. Indeed, since $z_{tr}\leq1$, $z_{eq}\leq1$ and $z_{tr}\geq z_{eq}$, then
\begin{equation}
\frac{1+z_{tr}}{1+z_{eq}}>1\,,
\end{equation}
implying $\beta>1$ as stated because $\alpha$ cannot be negative definite by construction, giving the additional requirement $\alpha>0$. These facts influence the experimental outcomes that we are going to describe in the next sections and are compatible with theoretical priors over $z_{tr}$ and $z_{eq}$. Once the constraints over $z_{tr}$ are known, it is therefore licit to display exclusion plots where Eq. \eqref{eq:ansatz} is no longer valid, displaying the phase space where $z_{eq}$ does not satisfy  $z_{tr}\geq z_{eq}$.

\subsection{How to get constraints over $z_{tr}$ and $z_{eq}$}

Here, our goal is to come up with a reasonable recipe for finding model independent constraints on $z_{tr}$ and $z_{eq}$ in analogy to the  cosmographic procedure. When we state ``model-independent'' we mean without postulating $f(z)$ \emph{a priori}, albeit that a few conditions have been reasonably assumed such as homogeneity and isotropy. For the sake of clearness, the method is therefore not fully independent of the universe description. However, it is much less model dependent than postulating a dark energy function and fit it with data, making it more attractive and useful in order to check departures from the $\Lambda$CDM paradigm.

We quoted above the cosmographic procedure that provides constraints on the so-called cosmographic parameters. These represent the set of free terms to evaluate in cosmography \citep{cosmography1, cosmography2, brah1,oluongo1,oluongo2-1,oluongo2-2,oluongo2-3,oluongo2-4}. This treatment consists in expanding the scale factor $a(t)$ around the present day, namely $t_0$ and in relating this expansion to observable quantities. To better focus on it, we therefore start with
\begin{equation}
a(t)=1+\sum_{n=1}^{\infty}\dfrac{1}{n!}\dfrac{d^n a}{dt^n}\bigg | _{t=t_0}(t-t_0)^n\,,
\label{eq:scale factor}
\end{equation}
and define the  cosmographic series, that for our purposes can be truncated up to the snap parameter, corresponding to fourth order in $(t-t_0)$. The jerk and snap parameters are given by
\begin{align}
j \equiv \dfrac{1}{aH^3}\dfrac{d^3a}{dt^3}\,, \qquad s \equiv \dfrac{1}{aH^4}\dfrac{d^4a}{dt^4}\,, \label{eq:j&s}
\end{align}
and writing down the corresponding series up to the fourth order in the cosmic time difference $t-t_0$ as follows
\begin{align}\label{serie1a}
a(t) =  1& - H_0 \Delta t - \frac{q_0}{2} H_0^2 \Delta t^2 +\nonumber\\
&-\frac{j_0}{6} H_0^3 \Delta t^3 + \frac{s_0}{24} H_0^4 \Delta t^4 + \ldots\,,
\end{align}
we can relate the cosmographic set\footnote{All the terms, including jerk and snap are evaluated at our time. This reflects the subscript ``0'' that we wrote in Eq. \eqref{serie1a}.}  to $\{z_{tr},z_{eq}\}$.

Clearly, there is an intrinsic difficulty to work with  $z_{eq}$ instead of $z_{tr}$. In fact, $q(z_{eq})\neq0$, while we notice that plugging the condition \eqref{semov} into cosmography  would reduce the complexity of our procedure.

In this paper, as we will stress also later in the text, we therefore directly fit the  transition redshift using the cosmographic method and then we exclude the regions in which $z_{eq}$ is forbidden. Clearly this represents our choice, essentially based on reducing the overall complexity of our fits by virtue of condition \eqref{semov}.

Thus, for the sake of completeness our main target
is not to constrain theoretical models, but rather to bound $z_{tr}$ from fits and exclude $z_{eq}$ consequently, as model independent as possible. Since this would represent a ``top-down'' approach, attempting to deduce the  kinematics directly from observations\footnote{Differently from a ``bottom-up'' approach, where one assumes
the dynamics of a given model postulated \emph{a priori}.}, we will be also able to test the validity of the standard paradigm \emph{a posteriori}.

Last but not least, for completeness the expansions of the Hubble rate,  luminosity and angular distances in terms of standard cosmography are reported in Appendix A.

It is now convenient to follow the steps summarized in the next section to frame the theoretical setup that we will follow throughout this manuscript.

\section{Theoretical setup}\label{sez3}

In this section, our concern is to determine the transition redshift constraints in a model-independent way. As we stressed above, we directly work with $z_{tr}$ to reduce the complexity by virtue of Eq. \eqref{semov} and exclude the regions in which $z_{eq}$ is forbidden \emph{a posteriori}. To do so, we intend to follow two distinct strategies, based on the following basic demands.

\begin{itemize}
\item[{\bf I.}] All our formal expansions are built up through cosmological quantities of interest that we directly compare with data, {\it i.e.} the Hubble rate, luminosity distance,  angular distance, and so forth. The expansions are computed around $z_{tr}$, with $q(z_{tr})=0$.
\item[{\bf II.}] Every contribution does not threaten to blow up, ruining the stability of the expansion, by fixing appropriate orders of Taylor series. In this respect, we evaluate the new cosmographic series where the cosmographic coefficients become functions of $z_{tr}$.
\item[{\bf III.}] We finally get constraints over $z_{eq}$, portraying formal \emph{exclusion plots} in which we report the regions where $z_{eq}$ can span by virtue of Eq. \eqref{eq:ansatz}.
\end{itemize}
Differently of the standard cosmographic set, namely $H_0,q_0, j_0$ and $s_0$ the corresponding priors are not known \emph{a priori}. So, the possible price we pay is that, after expanded in Taylor series, the new observable quantities of interest could be fully unbounded. To this end, an intriguing strategy to get hints toward the priors to use, is to somehow furnish a correspondence $j_0=j_0(z_{tr})$ and $s_0=s_0(z_{tr})$. This is the philosophy of the second method that we are going to describe below.
Henceforth we conventionally baptize the first strategy as \emph{direct Hubble expansion} (DHE), whereas the second as \emph{direct deceleration parameter expansion} (DDPE). Following the above considerations, the first treatment concerns the \emph{direct expansion} around $z=z_{tr}$ of the quantities of our interest, while the second requires the deceleration parameter is expanded around $z=z_{tr}$ and then it \emph{directly confronts} the original cosmographic sets with the new one. We describe in detail below the two approaches.

\subsection{First procedure: DHE method}

\noindent We first discuss the direct procedure of expanding around $z=z_{tr}$ the  Hubble rate. We immediately get
\begin{align}\label{primometodoHespanso}
H\,=\,&H(z_{tr})+H_{zt}^\prime(z-z_{tr})+\frac{1}{2}H_{zt}^{''}(z-z_{tr})^2+\nonumber\\
&\frac{1}{6}H_{zt}^{'''}(z-z_{tr})^3
+\mathcal O\Big[(z-z_{tr})^4\Big]\,.
\end{align}

The above expression should be  normalized to $H(z=0)\equiv H_0$  at $z=0$. At this stage, this additional constraint means that our new cosmographic series rescales all the other coefficients in terms of one of them, {\it i.e.}, the Hubble rate today. Indeed, choosing to expand $H$ to a given order $n$, it is arguable that only $n-1$ cosmographic coefficients are effectively independent.

To see this more clearly, we write the connection between the cosmographic series and the Hubble rate \citep{cosmography1}
\begin{subequations}\label{Hpunto}
\begin{align}
\dot{H} &= -H^2 (1 + q)\,,\\
\ddot{H}  &= H^3 (j + 3q + 2)\,,\\
\dddot{H}  &= H^4 \left [ s - 4j - 3q (q + 4) - 6 \right]\,.
\end{align}
\end{subequations}
Now, we can use the identity
\begin{equation}
\frac{dz}{dt}\equiv-(1+z)H(z)\,,
\end{equation}
and so we immediately get the derivatives of the deceleration parameter with respect to $z$
\begin{subequations}\label{algebra2}
\begin{align}
\dfrac{dq}{dz}&=\dfrac{j-2q^2-q}{1+z}\,.\\
\frac{d^2q}{dz^2}&=-\frac{1}{1+z}\left[2\frac{dq}{dz}(2q+1)-\frac{dj}{dz}\right]\,,\label{syg1}
\end{align}
\end{subequations}
So, Eqs. \eqref{Hpunto} - \eqref{algebra2} certify that only $n-1$ coefficients are really independent. At this stage, since $a\equiv(1+z)^{-1}$, with $a(z=0)=a_0=1$, we get from Eqs. \eqref{Hpunto} the following expressions
\begin{subequations}\label{Hprimi}
\begin{align}
H^{'''}  &=H\frac{3q^2(1+q)-j(3+4q)-s}{(1+z)^3}\,,\\
H^{''}  &= 
\frac{H(j-q^2)}{(1+z)^2}\,,\\
H^\prime &= \frac{H(1+q)}{1+z}\,,
\end{align}
\end{subequations}
that at the transition time reduce to
\begin{subequations}
\begin{align}
H^{'''}_{tr}&=-\frac{H_{tr}(3j_{tr} + s_{tr})}{(1+z_{tr})^3}\,,\\
H^{''}_{tr}&=\frac{H_{tr}j_{tr}}{(1+z_{tr})^2}\,,\\
H^\prime_{tr}&=\frac{H_{tr}}{1+z_{tr}}\,.
\end{align}
\end{subequations}
As additional constraint over Eqs. \eqref{Hprimi}, we plug $z=0$ in Eq. \eqref{primometodoHespanso} and  require $H(z=0)\equiv H_0$. This procedure further simplifies Eq. \eqref{primometodoHespanso} up to the selected order. In the case of orders three and two respectively we get

\begin{align}\label{1primometodoHespanso2}
&\mathcal E^{(3)}(z)\simeq\nonumber\\
&\frac{6 + z (6 + 3 j_{tr} z - (3 j_{tr} + s_{tr}) z^2) + 12 z_{tr} }{6 +
 z_{tr} (12 + z_{tr} (6 + s_{tr} z_{tr} + j_{tr} (3 + 6 z_{tr})))}\,\,+\nonumber\\
 &   \frac{3 z (4 + s_{tr} z + j_{tr} (-2 + 4 z)) z_{tr}}{6 +
 z_{tr} (12 + z_{tr} (6 + s_{tr} z_{tr} + j_{tr} (3 + 6 z_{tr})))}\,\,+\\
 &  \frac{3 (2 + j_{tr} + (2 - 5 j_{tr} - s_{tr}) z) z_{tr}^2 + (6 j_{tr} + s_{tr}) z_{tr}^3}{6 +
 z_{tr} (12 + z_{tr} (6 + s_{tr} z_{tr} + j_{tr} (3 + 6 z_{tr})))}\,,\nonumber
\end{align}

\noindent and

\begin{equation}\label{2primometodoHespanso2}
\mathcal E^{(2)}(z)=H_0 \left[\frac{2 + j_{tr} z^2 + 2 z (1 + z_{tr} - j_{tr} z_{tr}) + z_{tr} (2 + j_{tr} z_{tr})}{2 +
 z_{tr} (2 + j_{tr} z_{tr})}\right]\,.
\end{equation}
Thus, we set the following constraints respectively for Eqs. \eqref{1primometodoHespanso2} and \eqref{2primometodoHespanso2}

\begin{subequations}\label{constraintss}
\begin{align}
H_{tr} &= \frac{6 H_0 (1 + z_{tr})^3}{
 6 + z_{tr} (12 + z_{tr} (6 + s_{tr} z_{tr} + j_{tr} (3 + 6 z_{tr})))}\,,\\
H_{tr} &= \frac{2 H_0 (1 + z_{tr})^2}{ 2 + z_{tr} (2 + j_{tr} z_{tr})}\,,
\end{align}
\end{subequations}

with the obvious requirement  that
$j_0\neq j_{tr}$ and
 $s_0\neq s_{tr}$, besides $q_{tr}=0$.
However, since we are limiting our analysis to  late-time, the corresponding redshifts are confined around $z\lesssim 2$. Within this sphere, we can approximately  assume any further orders beyond snap to be negligible.
Immediately, from Eqs. \eqref{1primometodoHespanso2} and \eqref{2primometodoHespanso2}, we notice new priors on $j_{tr}$ and $s_{tr}$ are needful, as we will discuss later in the text.

For the sake of completeness, it is evident the denominators of Eqs. \eqref{1primometodoHespanso2} and \eqref{2primometodoHespanso2} do not limit the priors to impose, although  $H_0$ would do, as due to the $H_0$ tension.

Finally, it is remarkable to stress a strong multiplicative degeneracy in fitting $z_{tr}$ with $j_{tr}, s_{tr}$. In particular, this would imply a further uncertainty in the fitting procedures.

\subsection{Second procedure: DDPE method}

The second procedure, as stated above, consists of expanding the deceleration parameter in a Taylor series up to a given order around the transition redshift $z_{tr}$. For practical reasons, we fixed two hierarchical orders in analogy to the DHE, {\it i.e.}, the first and second orders of Taylor expansions of the deceleration parameter as
\begin{subequations}\label{eq:exp q}
\begin{align}
q^{(I)}&\simeq q(z_{tr})+\frac{dq}{dz}\bigg|_{z_{tr}}(z-z_{tr})
\,,\\
q^{(II)}&\simeq q^{(I)}+\frac{1}{2}\frac{d^2q}{dz^2}\bigg|_{z_{tr}}(z-z_{tr})^2
\,.
\end{align}
\end{subequations}
By definition, $q(z_{tr})$ vanishes and in the redshift interval $z<1$ we can approximate the value of the deceleration parameter to the value inferred from cosmography today, $q_0$, by simply taking $z=0$ inside Eqs. \eqref{eq:exp q}.
One can therefore compute the extreme values of $q$ within the interval $0<z<z_{tr}$, corresponding to a minimum and maximum value of $q$. In particular, the deceleration parameter is negative definite to guarantee current acceleration. It follows that
\begin{eqnarray}
\left\{
  \begin{array}{ll}
   q_{0,min}=-\dfrac{dq}{dz}\Big|_{z_{tr}}z_{tr}+\dfrac{1}{2}\dfrac{d^2q}{dz^2}\Big|_{z_{tr}}z_{tr}^{2}
   , & \qquad z=0\,, \\
   \,\\
     q_{0,max}=0\,, \ & \qquad z=z_{tr}\,,
  \end{array}
\right.
\end{eqnarray}
in agreement with the fact that the transition occurs when $q$ changes sign (passing from a positive to a negative value). The above request of minimum and maximum values for $q$ leads to small errors in expanding $q$ around $z=z_{tr}$.

The procedure to get the Hubble rate that we adopt here for our  fits is the following. We plug Eqs. \eqref{eq:exp q} inside the second Friedmann equation of Eqs. \eqref{fry}. Then, known $q$ we infer $H$ in function of $z_{tr}$ and we require $H(z=0)=H_0$ as we did before. It is remarkable to notice that taking the direct expansion of $H$ around $z=0$, as reported in Appendix A and
substituting Eqs. \eqref{eq:exp q} with $z=0$ would be also possible and apparently easier. However, the former strategy leads to worse results in computation. In particular, this technique is not reasonable because of the main caveat: taking a $z=0$ Hubble expansion and plugging and expanded $q$ with $z=0$ would induce \emph{an approximation within an approximation}. This double approximation is therefore a source of further errors in the numerical computations that we are going to present.  Motivated by these reasons, we decide to follow the other strategy to compute the quantities to fit. Hence, we immediately get for order two and three the following normalized Hubble rates

\begin{align}\label{Hmetodo2ordine2e3}
&\mathcal E^{(2)}(j_{tr},z_{tr})=\exp^{\frac{j_{tr} }{1 + z_{tr}}z} (1 + z)^{1 - j_{tr}}\,,\\
\,\nonumber\\
&\mathcal E^{(3)}(j_{tr},z_{tr})=
\mathcal E^{(2)}\exp^{\frac{(4 j_{tr} + s_{tr}) (z - 4 z_{tr}-2)z}{4 (1 + z_{tr})}}(1 + z)^{{1\over 2} (4 j_{tr} + s_{tr}) (1 + z_{tr})}\,.\nonumber
\end{align}

\begin{figure*}\label{figura}
\centering
\includegraphics[height=0.32\hsize,clip]{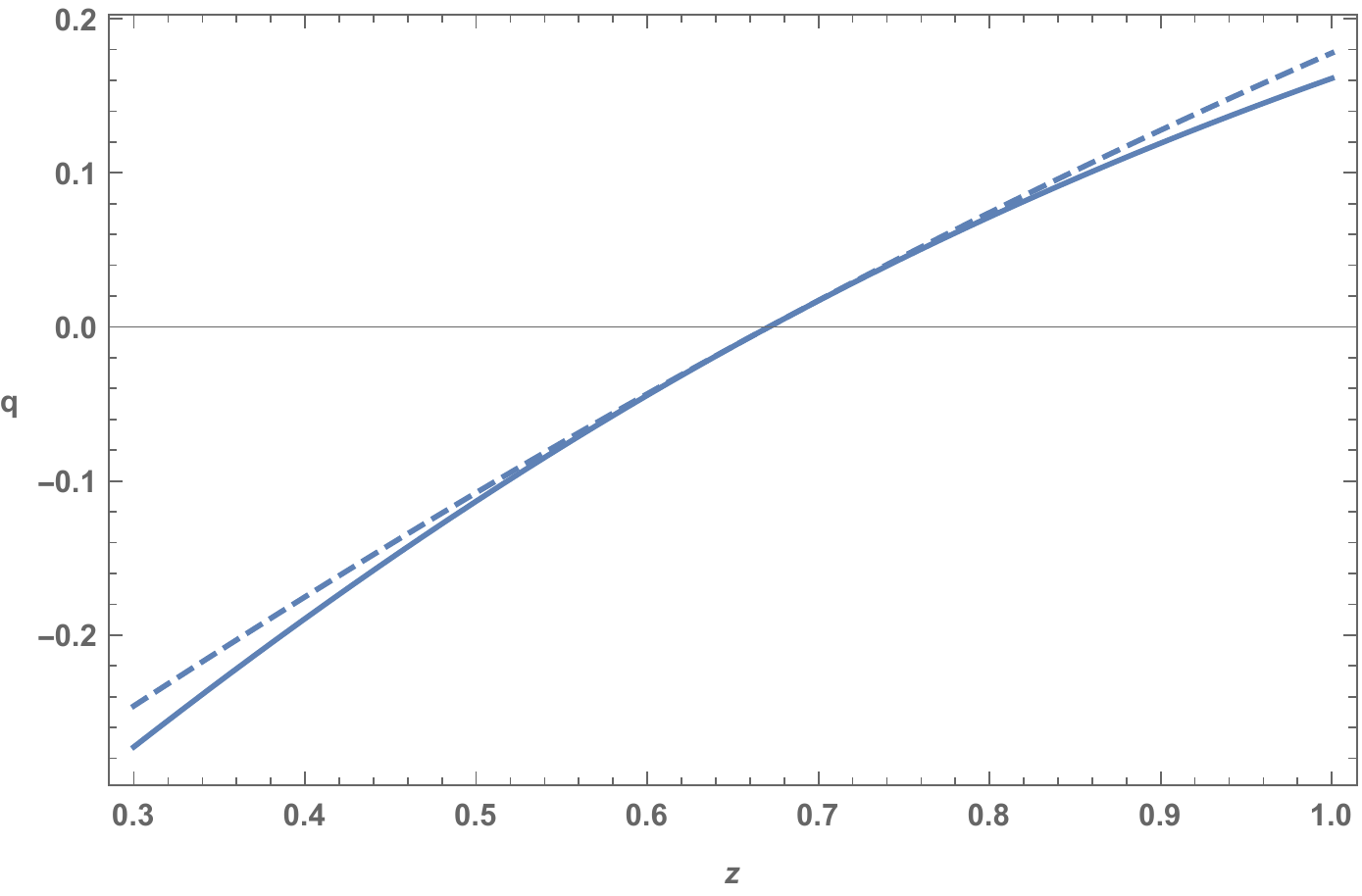}
\includegraphics[height=0.32\hsize,clip]{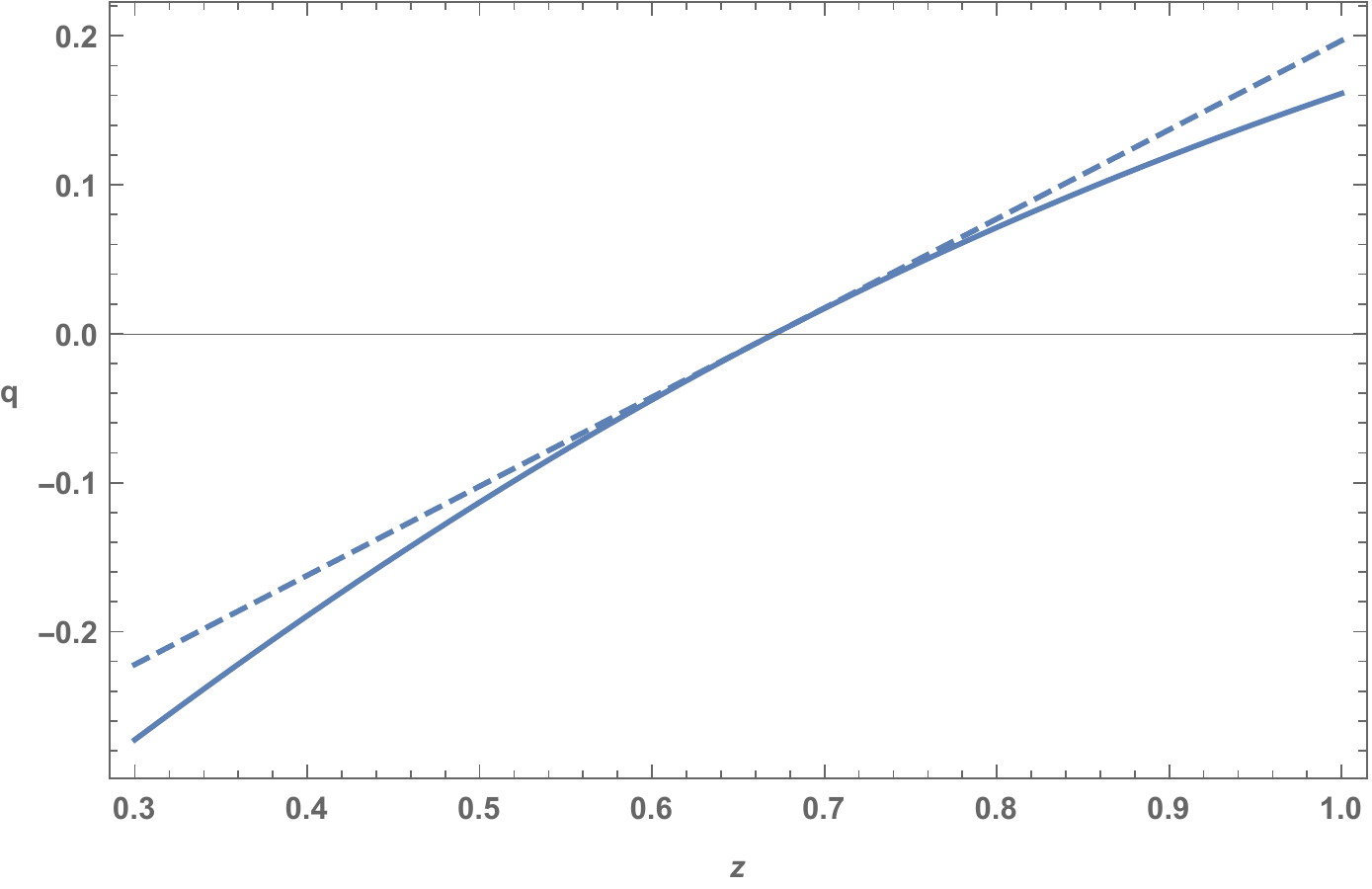}
\caption{Confront between $\mathcal N=2$ DHE and $\mathcal N=1$ DDPE approximations with the $\Lambda$CDM deceleration parameter within the interval $z\in[0.3;1]$. Here we compare the deceleration parameters up to the jerk term. The indicative values here used are  $\left\{z_{tr}, j_{tr}\right\}=\left\{0.7, 1\right\}$, whereas the indicative value for the mass is $\Omega_{m,0}=0.3$. In the intermediate region, {\it i.e.}, $z\in[0.5;0.9]$ the two approximations are practically indistinguishable, but as $z\rightarrow0$ the DHE method seems to better approximate the $\Lambda$CDM deceleration parameter. In the DHE method,  to compute the deceleration parameter, we combined the exact formula $q=-1+(1+z)\frac{dH}{dz}H^{-1}$ with $H(z)$ got from Eq. \eqref{2primometodoHespanso2}. }
\end{figure*}

\begin{figure*}\label{figura}
\centering
\includegraphics[height=0.32\hsize,clip]{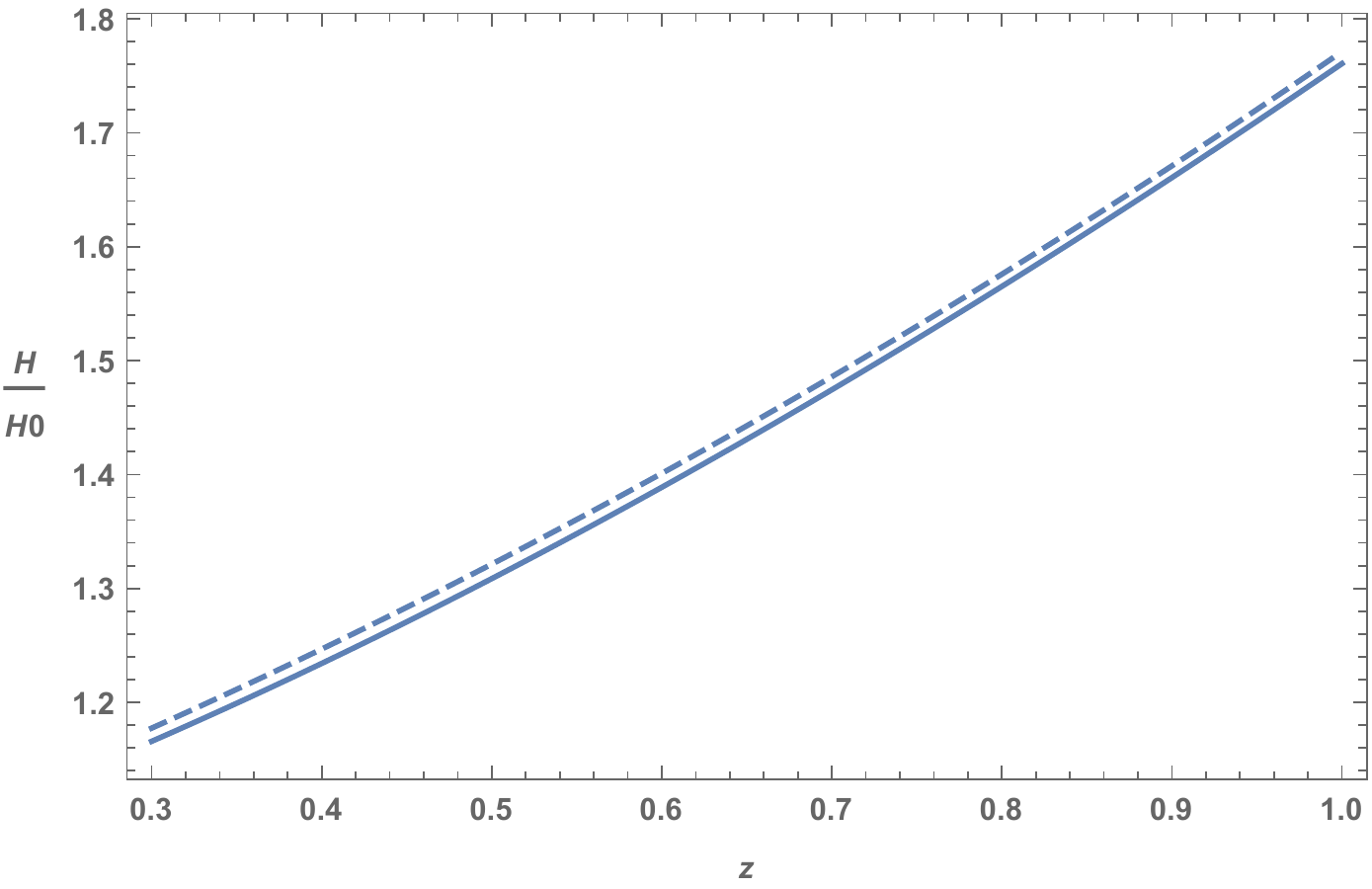}
\includegraphics[height=0.32\hsize,clip]{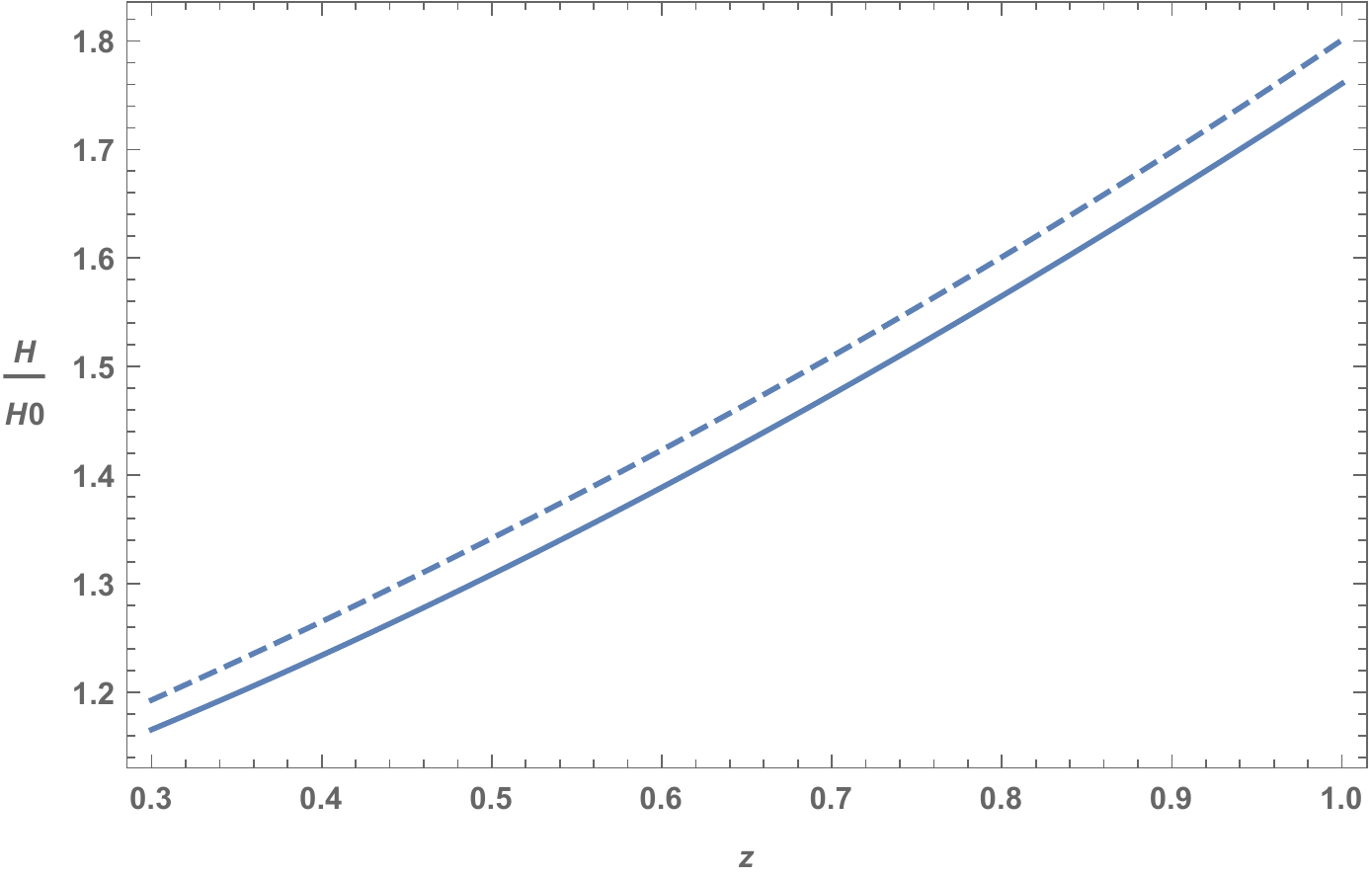}
\caption{Confront between $\mathcal N=2$ DHE and $\mathcal N=1$ DDPE approximations with the $\Lambda$CDM Hubble parameter within the interval $z\in[0.3;1]$. Here we compare the Hubble parameters up to the jerk term. The indicative values here used are  $\left\{z_{tr}, j_{tr}\right\}=\left\{0.7, 1\right\}$, whereas the indicative value for the mass is $\Omega_{m,0}=0.3$. In the intermediate region, {\it i.e.}, $z\in[0.5;0.9]$ the two approximations are practically indistinguishable for DHE, but similar in form for DDPE. Both approximate the $\Lambda$CDM Hubble parameter in terms of its shape.  }
\end{figure*}

\begin{figure*}
\centering
\includegraphics[height=0.29\hsize,clip]{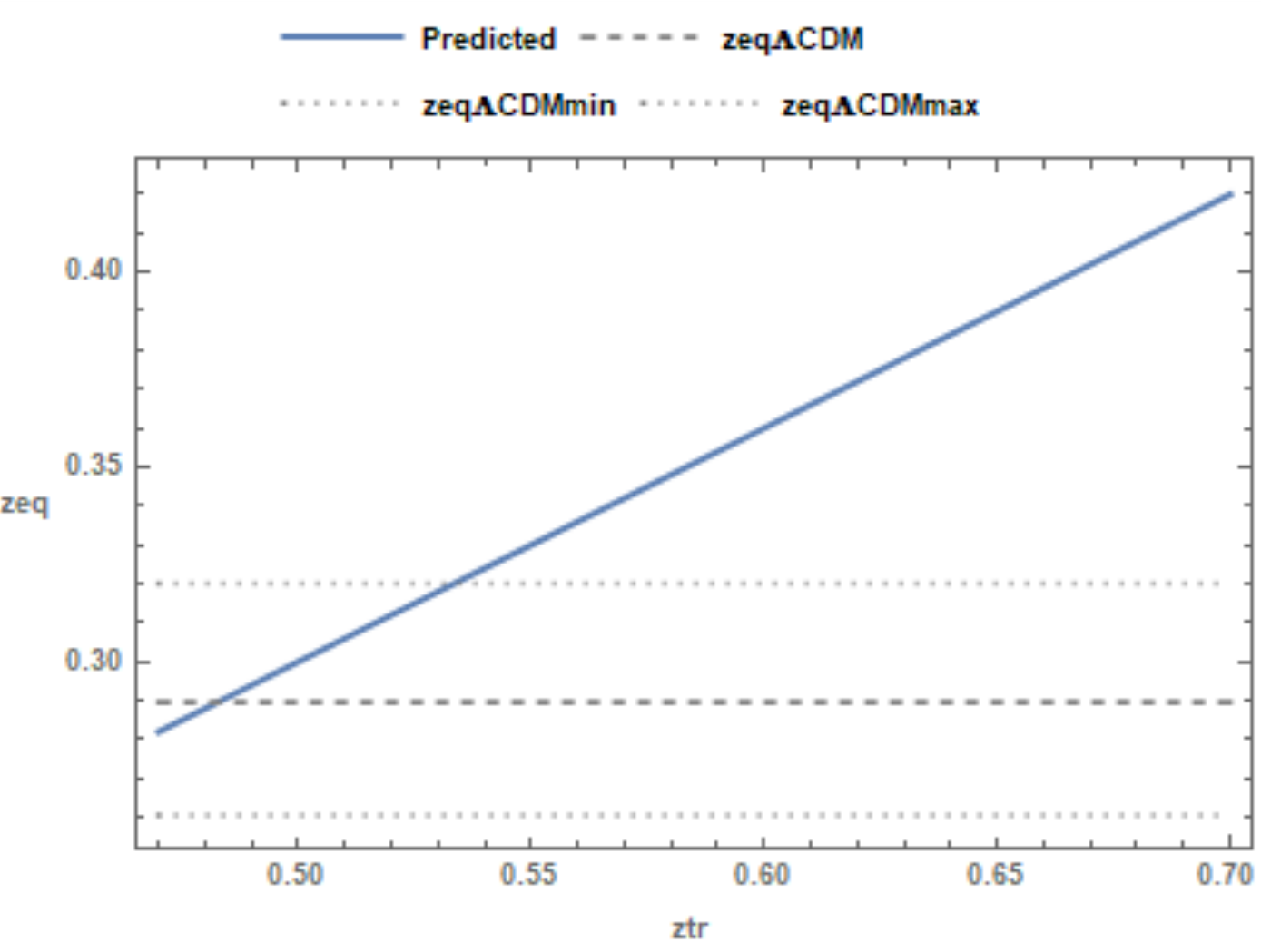}
\includegraphics[height=0.29\hsize,clip]{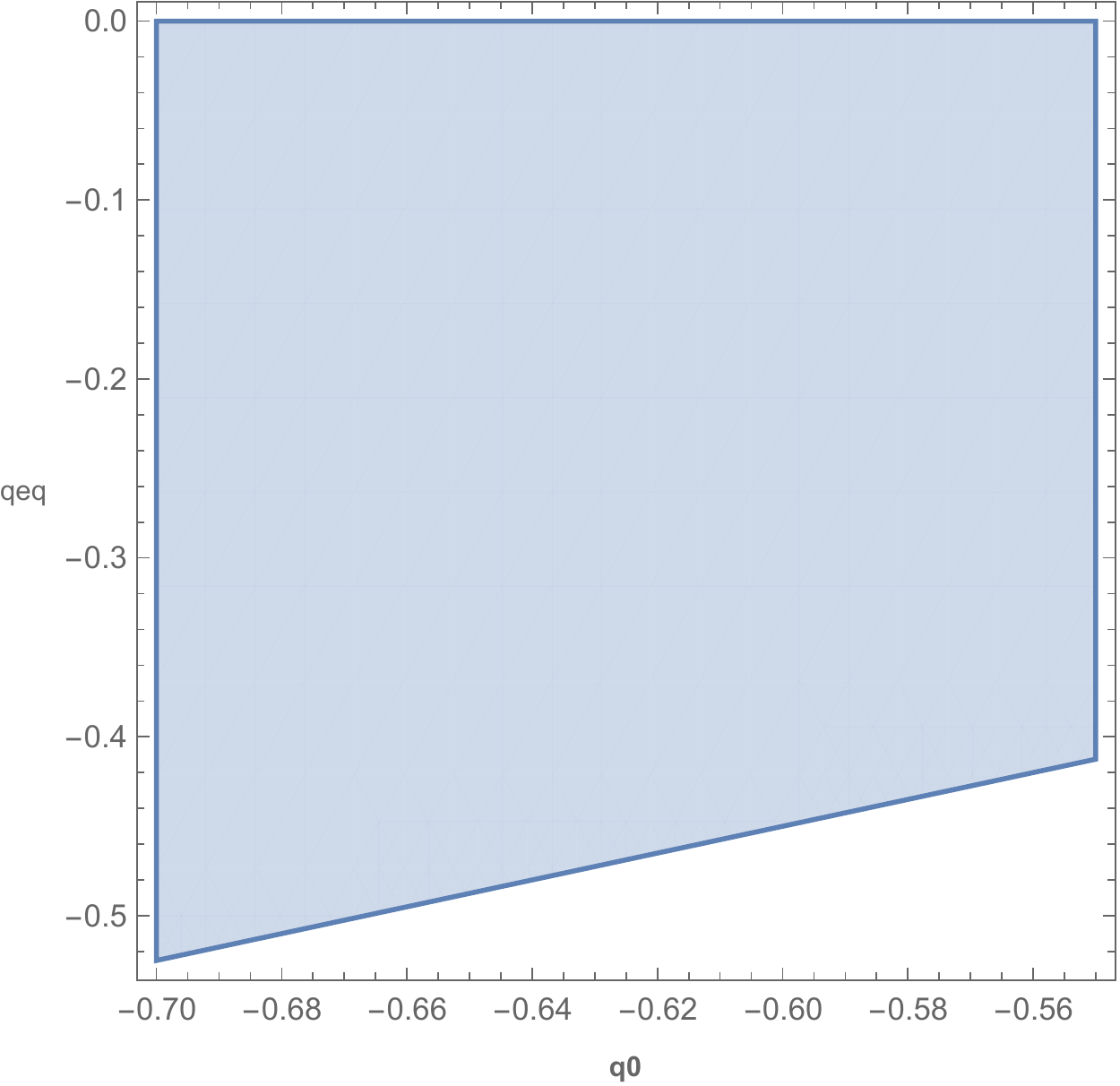}
\includegraphics[height=0.29\hsize,clip]{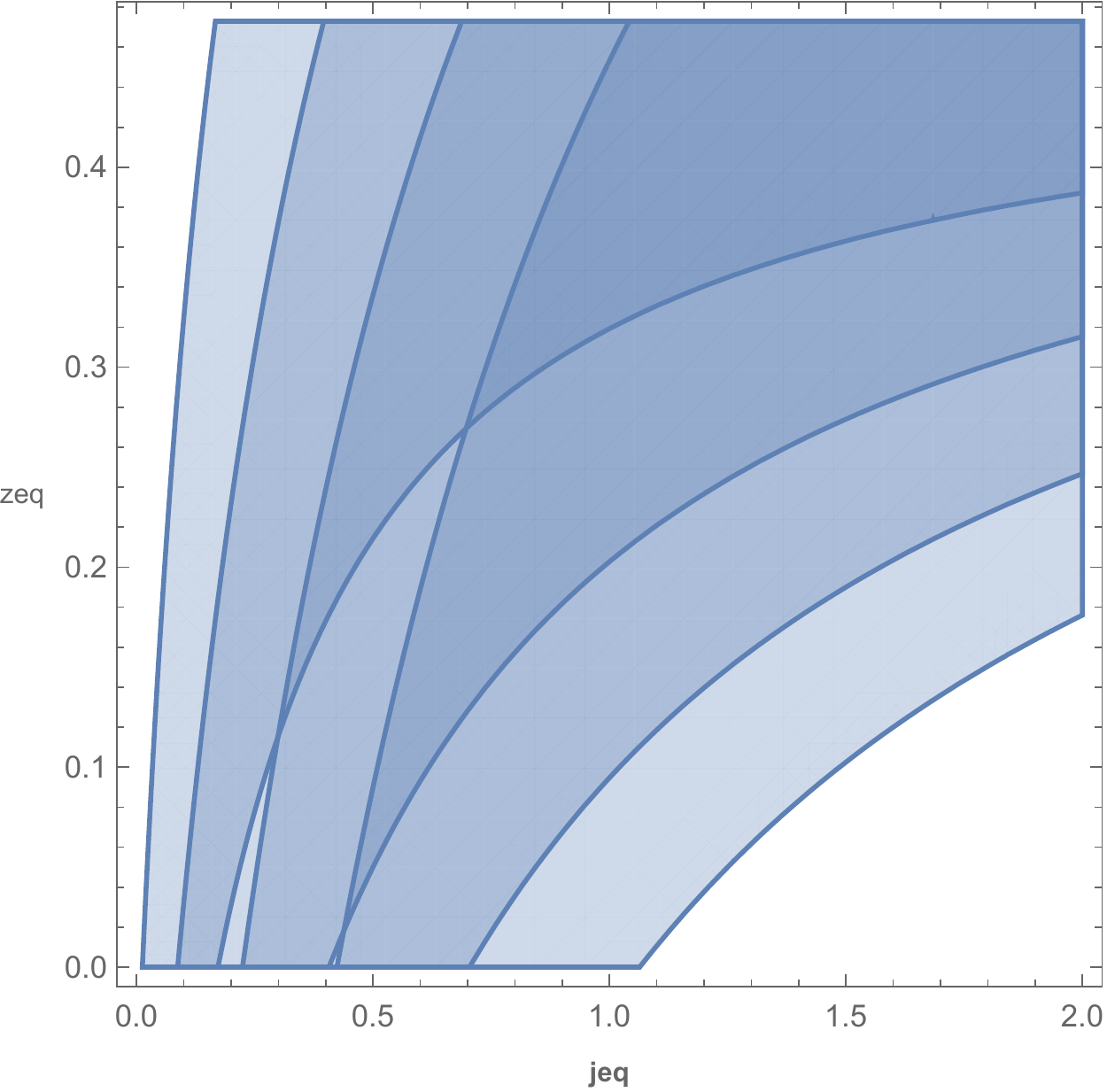}
\caption{Exclusion plots with the allowed regions for $z_{eq}$ and  comparison with the $\Lambda$CDM predictions. We here consider $z_{eq\Lambda CDM, min}$ and $z_{eq\Lambda CDM, min}$ using the Planck's  matter density, {\it i.e.}, $\Omega_{m,0}=0.318\pm0.015$. The minimum and maximum values are computed  considering $1\sigma$ errors. The figure on the left shows that the $\Lambda$CDM predictions are in a short interval only with respect to our approximations. However we notice $z_{eq}$ to lie in feasible intervals as $z_{tr}$ increases. The central figure indicates the expected values of $q_{eq}$ with respect to $q_0$. The white region is forbidden, indicating that $q_{eq}>q_0$ in the accessible region. The plot is got adopting ${q_0 - q_{eq}\over q_0}\in[0.25;1]$. The final plot on the right is conventionally made with $q_{eq}=\left\{-0.5;-0.375;-0.250;-0.125\right\}$ and shows that $j_{eq}\geq1$ is favorite to have $z_{eq}\geq0.3$, in fulfillment of the left plots, as one can see from the darker matched region. Concluding the most suitable intervals for $z_{eq}$ lead to $z_{eq}\in[0.2;0.45]$ for $z_{tr}\leq 0.7$ and $q_{eq}\geq -0.5$.}\label{altrafigura}
\end{figure*}

\section{Methodology}\label{metodo}

The strategy is to approximate distances making use of the standard formula
\begin{subequations}\label{basicapprox}
\begin{align}\label{base}
r_0\equiv\int_0^{z}\dfrac{dz^\prime}{H(z^\prime)}\,.
\end{align}
\end{subequations}
In particular, to get the luminosity and angular distances, as we report below, the idea is to integrate Eqs. \eqref{1primometodoHespanso2} and \eqref{2primometodoHespanso2} for the DHE method and Eqs. \eqref{Hmetodo2ordine2e3} for the DDPE method in order to fix the cosmographic coefficients evaluated around the transition time.

As above stated, we thus consider for our numerical purposes three observable quantities, {\it i.e.}, two distances, the luminosity and angular ones, and the Hubble rate, truncated with the above orders. For every quantity since we imposed a vanishing curvature parameter, say $\Omega_k=0$   \citep{Planck2018}, we do not care about the likely degeneracy between transition and $\Omega_k$. In what follows, we summarise the numerical strategy for each of the aforementioned quantities that we handle in our fits. Moreover, we report the main caveats that we can encounter throughout our computation.

\subsection{Overcoming bias the truncated series issues}

Expanding in terms of $z_{tr}$ with DHE and DDPE strategies guarantees to get constraints over the free parameters, namely $z_{tr}$ and  $H_{tr}, j_{tr}, s_{tr}$. Thereafter, one can make
comparisons to the measured quantities and accept or
reject the realization of our fits. The former could be due to the truncating series errors and/or to bias affecting the procedure itself.

In case one restarts the procedure with a new realization of the underlying  parameters, it is useful to understand how to reduce bias and systematics on our numerical fits, but also the correlations that occur among parameters, {\it i.e.}, intimately related to the degeneracy among coefficients\footnote{For example, in the spatially-flat $\Lambda$CDM model, one encounters a linear correlation between the snap parameter and $q_0$, namely $s_{0,\Lambda CDM}=-2-3q_{0,\Lambda CDM}$.}.  As the true cosmological bounds over transition is expected to marginally lie around $\Lambda$CDM predictions, we expect that a successful numerical set follows this trend in a small neighborhood of its best fits, indicating unreasonable degeneracy.

To overcome the above caveats and removing unwanted degeneracies as discussed above, we propose
a new route that uses Hubble's expansions as input functions inside Eq. \eqref{base}. In other words, instead of expanding Eq. \eqref{base} as one commonly does in standard cosmography, we take the above computed Hubble's expansions for the DHE and DDPE methods and evaluate the corresponding exact integrals from Eq. \eqref{base}.

Using this strategy, we see the dispersion and bias turn out to be much smaller than standard cosmography, speeding up \emph{de facto} the overall
numerical computations.

\subsection{The luminosity distance and $z_{tr}$}

We start with the standard definition for the luminosity distance in the case of spatially flat universe. We have
\begin{equation}
\label{dlHz2}
d_{\rm L}(z)=(1+z)\int_0^z\dfrac{dz'}{H(z')}=(1+z)r_0\,.
\end{equation}
In particular, Eq. \eqref{dlHz2} can be used in the following way: we construct the expanded Hubble parameter up to the order that we require. Then we insert the Hubble expansion within Eq. \eqref{dlHz2}, depending on the transition redshift and the cosmographic series evaluated at the transition. This strategy makes it position to understand the role played by the two coefficients $j_{tr}$ and $s_{tr}$, with analogous degeneracy problems than standard cosmography.

Another intriguing fact to remark is that it could be possible to directly expand Eq. \eqref{dlHz2}, instead of inserting a Hubble expansion inside it. However, we follow the strategy presented in \citep{oluongo1}, in which it has been argued that the cosmographic expansions with the aforementioned assumptions do not provide a significant  increase of degeneracy among cosmographic coefficients, enabling the one-to-one identification:
\begin{equation}\label{serievera}
d_L\left(z;\Omega_{m,0};\Omega_X\right)\rightarrow d_L\left(z;z_{tr};\theta\right)\,,
\end{equation}
where $\theta$ represents the cosmographic set from which $d_L$ is thought to depend on, while $\Omega_X$ is the generic dark energy density.

The convergence of truncated Hubble rate at a given order jeopardizes the overall analysis and may produce systematics in our computation. To expect a more stable numerical output from our fits, we leave our analyses fully free to vary in the priors. In fact, as jerk and snap parameters change, the transition redshift can severely switch, suggesting that fixing cosmographic parameters in our analyses is not suitable to work with. The marginalization procedure is also dangerous since we do not know \emph{a priori} how much priors could influence our analysis. In fact, even the choice of priors has been as wide as possible, within a range that possibly does not influence the analysis itself.

It is remarkable that dark energy density does not  modify the expected bounds over $z_{tr}$ and $z_{eq}$, because the strategy here involved is fully model independent, besides the choice of zero spatial curvature. The only limitation could be to consider that well-motivated dark energy models do not predict transition outside the sphere $z\leq 1$. So, under the hypothesis of barotropic dark energy paradigms, we assume to rule out the models that show fits outside the sphere $z\leq1$.  More complicated cases are however possible, but weakly supported by other observations. For example, assuming modified and/or extended theories of gravity, providing highly different dark energy model, is disfavoured by current bounds in the sphere $z\leq1$. The Occam razor suggests us to work out at late times, the simplest barotropic dark energy models, instead of invoking more complicated paradigms that would change the series we are handling and modify the expected ranges for the transition redshift.

\subsection{The angular distance and $z_{tr}$}

In this case, we start with the standard definition got from BAO, where the angular distance can be described in terms of uncorrelated and correlated data points. We write the distance
\begin{equation}
\label{eq:DV}
d_{\rm z}({\bf x},z) \equiv r_{\rm s}\left[\frac{H({\bf x},z)}{cz}\frac{\left(1+z\right)^2}{d_{\rm L}^2({\bf x},z)}\right]^\frac{1}{3}\,,
\end{equation}
with
\begin{equation}
A_{\rm z}({\bf x},z) \equiv \frac{\sqrt{\Omega_{\rm m}}H_0 r_{\rm s}}{cz d_{\rm z}({\bf x},z)}\,,
\end{equation}
where the comoving sound horizon,  $r_{\rm s}$, depends on baryon drag redshift $z_\text{d}$ and ${\bf x}$ is the set of free parameters involved into calculations.

In what follows, we only focus on uncorrelated BAO data pints. This choice reduces the data dependence of our fits. Finally, we do not use BAO catalogs alone, but only with SNe Ia and Hubble data to guarantee the less predicted biases from simulations and the robustness of our fits.

\subsection{The observational Hubble expansions}

To concern with OHD data, we  directly expand the Hubble rate as we previously highlighted for the DHE and DDPE methods. Every  $H(z)$ expansion formally provides
\begin{eqnarray*}\label{ansatz3}
    H(q_0,j_0,s_0)\rightarrow H(z,z_{tr},j_{tr},s_{tr})\,,
\end{eqnarray*}
and then, we take the expansions of $d_L$ and $d_A$ and plug the new definitions of $H$ expanded in the two forms, for the DHE and DDPE methods. Once the functions are got, we experimentally fit our truncated models directly with data.

We combine the three data sets, SNe Ia, BAO and OHD, to get limits over the transition redshift. However, our first fits are prompted using OHD alone. Using a single data set with not so much data points  clearly compromises the overall accuracy.
For these reasons, for all our computations, we need to study different sets of
parameters with fixed orders, leading to a hierarchy in our numerical analyses.

\subsection{Getting limits on the equivalence redshift}

Here we are interested in understanding how to obtain constraints on $z_{eq}$ starting from our previous prescriptions on $z_{tr}$ presented above. To do so, notice that it is possible to expand $q(z)$ around $z_{eq}$, in analogy to what we computed around $z\simeq z_{tr}$. Thus,  at first order around $z_{eq}$ we get
\begin{equation}\label{eq:exp qzeq}
q=q(z_{eq})+\frac{dq}{dz}\bigg|_{z_{eq}}(z-z_{eq})\,.
\end{equation}

In analogy to the previous calculations, substituting $z=z_{tr}$ and $z=0$ into Eq. \eqref{eq:exp qzeq}, we have

\begin{subequations}\label{eq:exp q3}
\begin{align}
q(z_{tr})&=q(z_{eq})+\frac{dq}{dz}\bigg|_{z_{eq}}\left(z_{tr}-z_{eq}\right)=0\,,\label{laprimadieq:exp_q3}\,\\
q(0)&=q(z_{eq})-\frac{dq}{dz}\bigg|_{z_{eq}}z_{eq}=q_0\,.\label{lasecondadieq:exp_q3}
\end{align}
\end{subequations}

Our intend is to get an expression that relates $z_{tr}$ with $z_{eq}$. At a first glance, we can first evaluate, from Eq. \eqref{laprimadieq:exp_q3}, the quantity $\frac{dq}{dz}\big|_{z_{eq}}=\frac{q_{eq}}{z_{eq}-z_{tr}}$ and then we  plugg it into Eq. \eqref{lasecondadieq:exp_q3} to get

\begin{equation}\label{q0qeq}
z_{eq} = \frac{q_0 - q_{eq}}{q_0} z_{tr}\,.
\end{equation}
The expression in Eq. \eqref{q0qeq} relates the present value of the deceleration parameter with the equivalence and transition redshifts. In principle, for fixed $q_0$, knowing the intervals in which $q_{eq}$ can span, it is possible to infer $z_{eq}$, once we measured the allowed values for $z_{tr}$,  portrayed in Tab. \ref{tabe}.

In addition, it is straightforward to  notice that the following equality holds
\begin{equation}
\frac{dq}{dz}\bigg|_{z_{eq}}=\frac{j_{eq}-2q_{eq}^2-q_{eq}}{1+z_{eq}}=\frac{q_{eq}}{z_{eq}-z_{tr}}\,,
\end{equation}
so that, solving with respect to  $z_{tr}$ and collecting in terms of $z_{eq}$, we write
\begin{equation}\label{ztrzeqconfronto}
z_{tr}\,=\,-\frac{q_{eq}}{j_{eq} - q_{eq} (1 + 2 q_{eq})} + \frac{j_{eq} - 2 q_{eq} (1 + q_{eq})}{j_{eq} - q_{eq} (1 + 2 q_{eq})}z_{eq}\,.
\end{equation}
Thus, confronting Eqs. \eqref{eq:ansatz} and \eqref{ztrzeqconfronto}, we infer
\begin{subequations}\label{alfabeta}
\begin{align}
\label{43a}
\alpha & = -\frac{q_{eq}}{j_{eq} - q_{eq} (1 + 2 q_{eq})}\,,\\
\,\nonumber\\
\label{43b}
\beta & =\frac{j_{eq} - 2 q_{eq} (1 + q_{eq})}{j_{eq} - q_{eq} (1 + 2 q_{eq})}\,.
\end{align}
\end{subequations}
By virtue of the considerations discussed in Sec. \ref{IIA}, recalling $\alpha>0$ and $\beta>1$, we stress that the condition $\alpha>0$ implies from Eq. \eqref{43a}
\begin{equation}\label{A7}
j_{eq}>q_{eq}(1+2q_{eq})\,,
\end{equation}
since we took $q_{eq}<0$ as expected\footnote{By construction, in the redshift domain in which dark energy dominates, the deceleration parameter is negative and as the redshift increases tends to become larger. Thus, since $z_{tr}>z_{eq}$, we have $q_{eq}\,<\,q_{tr}=0$.}. On the other hand, the condition $\beta>1$ is verified if the inequality $j_{eq} - 2 q_{eq} (1 + q_{eq})>j_{eq} - q_{eq} (1 + 2 q_{eq})$ holds in Eq. \eqref{43b}. However, since $j_{eq} - q_{eq} (1 + 2 q_{eq})>0$ because of Eq. \eqref{A7}, we therefore get a more stringent requirement over $j_{eq}$ from Eq. \eqref{43b}
\begin{equation}\label{A8}
    j_{eq}>2q_{eq}(1+q_{eq})\,,
\end{equation}
from which we argue the basic constraint on $q_{eq}$
\begin{equation}\label{conditio}
    -1<q_{eq}\leq -\frac{1}{2}\,.
\end{equation}
In particular, the left part of the above equality, namely $q_{eq}>-1$, derives from Eq. \eqref{A8}, and it is quite obvious because $q_{eq}$ cannot exceed a de Sitter phase, say $q_{dS}=-1$, for describing the dark energy evolution. The right part of the equality, namely $q_{eq}\leq -{1\over2}$, derives from both Eq. \eqref{A7} in which we made the assumptions $q_{eq}<0$ and $j_{eq}>0$. While the first of the former considerations has been previously discussed, the second one, on the contrary, is a natural request that one assume in cosmography \citep{aggiunta1}. In fact, the jerk parameter is thought to be positive-definite to guarantee the deceleration parameter to change its sign. Thus, there is no reason \emph{a priori} to imagine $j_{eq}$ negative by construction. Moreover, looking at Tab. \ref{tabe} we got all positive values of $j_{tr}$ and since $z_{tr}>z_{eq}$, though $z_{tr}\simeq z_{eq}$, it is improbable to get a negative value for $j_{eq}$ that rapidly jumps to a positive one at the transition.

Last but not least, since  $q_{eq}>q_0$ is naturally fulfilled to guarantee the universe speeds up at current time, using Eq. \eqref{q0qeq}, we immediately get a likely \emph{exclusion plot} for $z_{eq}$, employing the following bounds over $q_0, z_{tr}$ and $q_{eq}$
\begin{subequations}\label{intervalli}
\begin{align}
    q_0&\in[-0.7;-0.55]\,,\\
    z_{tr}&\in[0.47;1.18]\,,\\
    q_{eq}&\in[-0.5;0]\,,
\end{align}
\end{subequations}
and guaranteeing $z_{tr}\geq z_{eq}$ as reported in previous sections.
Here, the first of the above intervals has been taken from Ref.  \citep{aggiunta1}, where we single out the Taylor fits on $q_0$. These limits are clearly suitable because compatible with the most recent bounds on $q_0$, see e.g. \citep{revluongo,cosmography2}. The second interval over $z_{tr}$ is naively got from our results, see Tab. \ref{tabe}, where we considered the smallest and largest $z_{tr}$ mean values got from our computations\footnote{For the sake of clearly, we could in principle take a mean $z_{tr}$ and then take the above interval using propagated error bars. However, the choice \eqref{intervalli}  enables us to get  larger domains for $z_{tr}$.}. The last interval corresponds to the largest interval in which $q_{eq}$ can span by virtue of Eqs.   \eqref{A7} - \eqref{A8} and \eqref{conditio}. In fact,  since $q(z_{eq})$ is larger than $q_0$, as the cosmic speed up has not start yet at the equivalence, we are forced to assume $q(z_{eq})>q_0$ because dark energy did not reach the time to dominate over matter. Since $q_0\in[-1;0]$, it is licit to presume $q(z_{eq})\in[-0.5;0]$. Hence, we baptize the $z_{eq}$ plot with the name  \emph{exclusion plot}. This plot emphasizes the regions where $z_{eq}$ is not allowed to vary by means of Eqs. \eqref{A7}, \eqref{A8} and \eqref{conditio}. Thus, we portray the exclusion plot in Figs. \ref{altrafigura}.

Two additional considerations on $z_{eq}$ are now needful to clarify the above treatment. The first is that it is more convenient to pass from Eq. \eqref{q0qeq} instead of Eq. \eqref{ztrzeqconfronto} to get $z_{eq}$ plots, because we do not know \emph{a priori} the value of $j_{eq}$ that is an extra parameter entering Eq. \eqref{ztrzeqconfronto}. The second consideration is noticing that it is  not convenient to directly fit $z_{eq}$ following the strategy above developed for $z_{tr}$. The reason purely lies on statistics, {\it i.e.}, using $z_{eq}$ directly in the cosmographic fits leads to a severe degeneracy with $q_{eq}, j_{eq}$, but also with $z_{tr}$. So that, it would be much more convenient to infer $z_{eq}$ bounds, once that the results over $z_{tr}$ are got from the computation. Similar considerations have been made also in previous sections, when we discussed to work with $z_{tr}$ in order to reduce complexity.

\begin{table*}
\begin{center}
\footnotesize
\setlength{\tabcolsep}{0.3em}
\renewcommand{\arraystretch}{2}
\begin{tabular}{c c c c c c c }
\hline
\hline
Model & $Datasets$ & $H_{0}$     & $z_{tr}$ & $j_{tr}$ & $s_{tr}$ & $\chi^2 $ \\
\hline
\hline
$DHE\,(\mathcal N=2)$ & SN+OHD+BAO & $74.879_{-3.340(5.831)}^{+0.510(1.198)} $  & $0.646_{-0.158(0.229)}^{+0.020(0.031)} $ & $2.544_{-0.047(0.070)}^{+0.500(0.847)}  $ & $--$ & $1128.97$\\ 
$DHE\,(\mathcal N=2)$  & OHD & $67.659_{-4.198(6.497)}^{+3.825(9.246)}$ & $0.659_{-0.124(0.359)}^{+0.371(0.941)}$ & $0.982_{-0.750(0.940)}^{+0.519(1.375)}$ & $--$ & $14.7377$\\ 
\hline
$DHE\,(\mathcal N=3)$ & SN+OHD+BAO & $93.187_{-0.922(1.757)}^{+0.767(1.600)}$  & $0.656_{-0.017(0.036)}^{+0.025(0.047)}$ & $2.263_{-0.249(0.331)}^{+0.083(0.109)}$ & $-4.148_{-0.680(0.821)}^{+1.981(2.621)}$ &  $1658.14$ \\ 
$DHE\,(\mathcal N=3)$  & OHD & $76.331_{-9.111(13.556)}^{+3.946(10.083)}$
 & $0.473_{-0.057(0.156)}^{+0.178(0.624)}$
 & $2.964_{-2.128(2.823)}^{+0.578(1.676)}$
 & $0.196_{-1.363(2.034)}^{+1.433(2.526)}$
 & $12.9656$\\ 
\hline
$DDPE\,(\mathcal N=1)$ & SN+OHD+BAO & $75.091_{-1.862(2.571)}^{+0.170(0.647)}$  & $0.860_{-0.146(0.173)}^{+0.013(0.289)}
 $ & $2.160_{-0.018(0.004)}^{+0.237(0.291)}$& $--$ & 1129.01\\ 
$DDPE\,(\mathcal N=1)$  & OHD & $66.534_{-3.151(4.277}^{+2.908(6.780)}$
  & $0.700_{-0.119(0.280)}^{+0.260(0.438)}$ & $0.830_{-0.635(0.957)}^{+0.303(0.854)}$ & $--$ & $15.0466$\\ 
\hline
$DDPE\,(\mathcal N=2)$ & SN+OHD+BAO & $75.092_{-1.001(1.393)}^{+0.242(0.767)}$  & $1.183_{-0.032(0.054)}^{+0.002(0.011)}$ & $1.173_{-0.100(0.144)}^{+0.011(0.036)}$& $-5.826_{-0.058(0.091)}^{+0.170(0.280)}$ & $1129.04$\\ 
$DDPE\,(\mathcal N=2)$  & OHD & $76.702_{-5.779(10.366)}^{+4.853(10.397)}$  & $0.489_{-0.063(0.147)}^{+0.105(0.198)}$ & $2.804_{-0.915(1.245)}^{+0.434(0.957)}$ & $-17.564_{-2.935(6.289)}^{+6.040(7.261)}$ & $1129.04$ \\ 
\hline
\hline
\end{tabular}
\caption{MCMC results at the 68\% (95\%) confidence level for our different cosmographic techniques from the combination of data sets.  $H_0$ values are expressed in km/s/Mpc. The corresponding $1\sigma$ and $2\sigma$ confidence levels are portrayed in Appendix C, where we report the contour plots of our analyses. The $\chi^2$ values are here not normalized. }
\label{tabe}
\end{center}
\end{table*}

\section{Statistical analyses}\label{sez4}

Our fits utilise low redshift data from different surveys. To perform the numerical procedures, we work out MCMC simulations sampled within the widest possible parameter space over the cosmographic coefficients. The strategy is to modify the freely available \texttt{Wolfram Mathematica} code reported in \citep{codice} that makes use of the widely adopted Metropolis-Hastings algorithm. We thus confront our luminosity distance expressed in terms of the new cosmographic set, based on $z_{tr}$ and $z_{eq}$, directly with data, explicitly reporting how to reduce the dependence on statistical distributions through the algorithm itself. The numerical procedure is characterised by minimizing the chi squares computed with different data sets. So, we indicate with ${\bf x}$ the best fit cosmological parameters. The set ${\bf x}$ clearly minimizes the total $\chi^2$, constructed by means of SNe Ia,  BAO and OHD data sets, written as
\begin{subequations}
\begin{align}
&{\rm Fit\,\,1:}\quad \chi^2_{\rm tot}=\chi^2_{\rm OHD}\,,\\
&{\rm Fit\,\,2:}\quad\chi^2_{\rm tot}=\chi^2_{\rm SN}+\chi^2_{\rm BAO}+\chi^2_{\rm OHD}\,.
\end{align}
\end{subequations}
In particular, we report below the way in which each $\chi^2$ is evaluated.

\begin{itemize}
\item[{\bf \underline{SNe Ia}}] In the case of SNe Ia, we computed the $\chi^2$ considering the most recent \textit{\emph{Pantheon sample}}. Still now, it turns out to be the largest  combined sample consisting of $1048$ SNe Ia. The SNe Ia data points span within the interval $0.01<z<2.3$ \citep{2018ApJ...859..101S} and enable to get the corresponding distance modulus by
\begin{equation}\label{modulo}
\mu_{\rm SN}=m_{\rm B}- \left(\mathcal{M}-\bar\alpha \mathcal{X}_1+\bar\beta \mathcal{C} -\Delta_{\rm M}-\Delta_{\rm B}\right)\ ,
\end{equation}
that consists in a parametric version of the distance modulus based on $m_{\rm B}$ and $\mathcal{M}$, respectively the $B$-band apparent and absolute magnitudes. In Eq. \eqref{modulo}, $\mathcal{X}_1$ and $\mathcal{C}$ represent the SN light-curve stretch and the colour factors, respectively whereas $\bar\alpha$, $\bar\beta$ and $\Delta_{\rm M}$  show the luminosity stretch and colour and distance correction. Moreover, $\Delta_{\rm B}$ is a distance correction. It is formulated considering the host galaxy mass containing SNe Ia and it is got by predicted biases inferred from simulations.

The error propagation is not affected by the fitting procedure since uncertainties of each SN do not depend on $\mathcal{M}$. The form of SN  $\chi^2$ easily becomes
\begin{equation}
\chi^2_{\rm SN}=\left(\Delta \mathbf{\mu}_{\rm SN}- \mathcal{M}\mathbf{1} \right)^{\rm T} \mathbf{C}^{-1}
\left(\Delta\mathbf{\mu}_{\rm SN}-\mathcal{M} \mathbf{1} \right)\,,
\end{equation}
where the module of the vector of residuals takes the form $\Delta\mu_{\rm SN}\equiv \mu_{\rm SN}-\mu_{\rm SN}^{\rm th}\left({\bf x},z_i\right)$, while $\mathbf{C}$ is the covariance matrix. There we statistical and systematic uncertainties on the light-curve parameters are provided. We assume a flat prior to remove $\mathcal{M}$ from our fitting procedure. This process of marginalisation leads to\footnote{Marginalising over $\alpha$ and $\beta$ has not been performed. Their contributions enter SN uncertainties.
} %
\begin{equation}
 \chi^2_{{\rm SN},\mathcal{M}} = a + \log \frac{e}{2 \pi} - \frac{b^2}{e}\,,
 \label{eqn:chimarg}
\end{equation}
where $a\equiv\Delta\vec{\mathbf{\mu} }_{\rm SN}^{T}\mathbf{C}^{-1}\Delta\vec{\mathbf{\mu} }_{\rm SN}$, $b\equiv\Delta\vec{\mathbf{\mu} }_{\rm SN}^{T}\mathbf{C}^{-1}\vec{\mathbf{1}} $, and $e \equiv
\vec{\mathbf{1}}^T\mathbf{C}^{-1} \vec{\mathbf{1}} $.

\item[{\bf \underline{BAO}}]

BAO waves are observed as a peak in the large-scale structure correlation function. This process is produced in the early universe by the sound wave propagation leading to an angular distance measure, providing a comoving volume variation $D_{\rm V}^3({\bf x},z)$ at a given $z$, implying the BAO observable for uncorrelated data $d_{\rm z}({\bf x},z)$ through\footnote{Differently from SNe Ia,  BAO data are more model dependent as comoving sound horizon $r_{\rm s}(z_\text{d})$  depends upon the baryon drag redshift $z_\text{d}$.
}
\begin{equation}
\label{eq:DV}
D_{\rm V}^3({\bf x},z) \equiv \frac{c\,z}{H({\bf x},z)}\left[\frac{d_{\rm L}({\bf x},z)}{1+z}\right]^2\ \ \,,\ \ d_{\rm z}({\bf x},z) \equiv \frac{r_{\rm s} (z_\text{d})}{D_{\rm V}({\bf x},z)}\,.
\end{equation}

The MCMC simulations are performed according to the results prompted in  \citep{Planck2018}, whose  best fit values are  $z_\text{d}=1059.62\pm0.31$ and $r_{\rm s}(z_\text{d})=147.41\pm0.30$, with the BAO points reported in Table~\ref{tab:BAO} in Appendix B, in which the correlated BAO from the WiggleZ data \citep{2011MNRAS.418.1707B} have been excluded. This guarantees that we do not assume observable quantities depending upon $\Omega_{\rm m,0}$. With this recipe we take
\begin{equation}
\chi^2_{\rm BAO}=\sum_{i=1}^{N_{\rm BAO}} \left[\frac{d_{\rm z,i}^{\rm obs}-d_{\rm z}^{\rm th}({\bf x},z_i)}{\sigma_{d_{\rm z,i}}}\right]^2\ .
\end{equation}

\item[{\bf \underline{OHD}}]

The Hubble points, namely the set got from the differential age method, are determined from reconstructing Hubble sets measured through spectroscopy \citep{ohd}. Measurements of the age difference take $\Delta t$ and $\Delta z$ of couples of passively evolving galaxies. The idea is that these galaxies are formed at the same time. Under this hypothesis, we assume $\Delta z/\Delta t\simeq \frac{dz}{dt}$, leading to an approximate expression for the Hubble rate, whose update sample consists till now of  31 OHD data, see e.g. \citep{datiohd}. In particular, data here depend on stellar metallicity
estimates, population synthesis models, progenitor biases, and
the presence of an underlying young component in massive and
passively evolving galaxies. This data set invokes severe systematics that fortunately are limited up to a $3\%$ error rate at intermediate redshifts. Thus, systematic errors, expected in our computations, are safely kept below the statistical ones.

The overall treatment  allows to obtain model-independent estimations of the Hubble rate, providing the idea of  cosmic chronometers that lead to
\begin{equation}
H_{obs}=-\dfrac{1}{(1+z)}{\left(\dfrac{dt}{dz}\right)}^{-1}\,,
\end{equation}
spanning withing the interval $0<z\lesssim3$ and showing the corresponding chi square
\begin{equation}
\chi_{OHD}=
\sum_{i=1}^{31}\left(\dfrac{H_{th}(z_i)-H_{obs}(z_i)}{\sigma_H^2}\right).
\end{equation}

\end{itemize}

%

\subsection{Error propagation and numerical outcomes}\label{sez5}

Unfortunately, our method presents a  number of limitations, essentially related to the kind of approximations made throughout the analysis. We briefly list such limitations below

\begin{itemize}
\item[{\bf 1.}] Direct low fit errors  due to the multiplicative degeneracy between $j_{tr}$ and $z_{tr}$, as consequence of Eq. \eqref{basicapprox}. A way out to face the issue is to marginalize over $j_{tr}$ within a given set of plausible values for it. This procedure, however, does not remove the degeneracy and consequently $z_{tr}$ values are influenced by $j_{tr}$.
\item[{\bf 2.}] Errors due to the truncated cosmographic Hubble series, {\it i.e.}, we do not know \emph{a priori} how truncating series influences the fitting error bars. For these reasons we expect statistical differences using distinct hierarchies.
\item[{\bf 3.}] The same as above, but for the deceleration parameter, {\it i.e.}, errors due to the truncated expansions of $q(z)$,  consequence of the fact that we truncate it at a given unknown order.
\end{itemize}

Summing up one can imagine to increase the error bars $\sigma_{\theta}$ considering the relation

\begin{equation}\label{enlarging}
\sigma_{\theta}=\sqrt{\sum_\mu\bar \sigma_{exp,\mu}^2+\sum_{\nu=1}^{3}\bar \sigma_{(\nu)}^{2}}\,,
\end{equation}
where $\sigma_{exp}$ represents the experimental error directly got from our fits and reported in Tabs. \ref{tabe} while $\sigma^{(\nu)}$ are the errors \emph{induced} by the aforementioned points raised above. Notice that $\bar \sigma\equiv \frac{\sigma}{N}$, where $N$ is the number of errors induced by the approximation. Following the treatment of assuming underestimated error bars of Ref. \citep{verde}, we notice enlarging them through Eq. \eqref{enlarging} does not change significantly our findings since the errors would increase of about $\sim 10\%$. We thus believe the overall procedure works fairly well and even enlarging the errors does not furnish more information in our analysis.

For our numerical fits, we adopt the priors that follow: $z_{tr}\in[0;1]$, $H_{0}/100 km Mpc^{-1}\in[0.5;1]$, $j_{tr}\in[0.5;5]$ and $s_{tr}\in[-25;25]$ for all our computations.

\section{Theoretical discussion and comparison with models}\label{sez6}

Our two strategies made use of Taylor expansions and hierarchy between coefficients. The overall merit of each method seems to be roughly equivalent. The contours provide similar results that do not depart significantly from each other, as reported in Appendix C. The DHE method involves $H(z)$ expansions, whereas the DDPE method considers expansions of $q(z)$. In particular, from the results of Tab. \ref{tabe} we argue the DHE fits indicate that at $1\sigma$ the transition redshifts are compatible with the standard cosmological model, although they leave open the possibility that dark energy is not under the form of a genuine cosmological constant.

The use of OHD data only with fits up to $s_{tr}$ is clearly disfavored  than combining the three data sets together, SN+BAO+OHD, up to $s_{tr}$. Concerning OHD alone, the best suite of cosmological results is for $\mathcal N=2$, whereas deviations are evident in the case $\mathcal N=3$, {\it i.e.} $z_{tr}$ seems to decrease. The reason is that OHD data points are only 31 and then it is more difficult to constrain a further parameter, namely $s_{tr}$. The prize to pay is enlarging the $H_0$ outcomes and decreasing $z_{tr}$. The Hubble rate is not so far from current bounds, being in overall agreement with \citep{Planck2018} and \citep{riess}, within the $1\sigma$ confidence level. The error bars, in fact, are here extremely larger for all the coefficients, for the same reasons we quoted before. Even though the corresponding fits are not ruled out, they are not as feasible as for $SN+OHD+BAO$.

The jerk parameter is not well constrained, because as one fits up to $s_{tr}$, it dramatically increases with respect to the ones got up to $j_{tr}$. When the hierarchy is up to $j_{tr}$, it seems to converge to $j_0\simeq1$ as the Hubble rate decreases today. This appears in agreement with the $\Lambda$CDM expectations.  This would indicate that the larger $H_0$ bounds lead to larger jerk constraints, besides $\mathcal N=1$ for the DDPE method with OHD.

These strong departures disagree with the $\Lambda$CDM predictions, indicating that dark energy may be  favored in framing out the universe dynamics, at least with this kind of approximation. However, again, the problem could lie on the small number of points inside the OHD catalog.

Adding SNe Ia and BAO, the situation seems to change. Here, the DDPE fitting procedure seems to be more stable than DHE. The error bars at $1\sigma$ confidence level are quite similar than DHE, but the matching with the concordance paradigm is more precise. The constraints over $H_0$ do not mostly favor the Planck expectations, since seem to be in agreement with \citep{riess}.

Besides the case $\mathcal N=3$ of the DHE method with SN+OHD+BAO data sets, combining more than one data set appears essential to argue more suitable intervals on $z_{tr}$, albeit the case $\mathcal N=2$ with the same three data sets of the DDPE method surprisingly favors larger $z_{tr}$. The jerk parameter is again quite large, except for the last case we have cited. Comparing the outcomes over $s_{tr}$, we notice it agrees with a negative value, but severely different than the standard cosmological paradigm. Concluding, even in this case dark energy is not excluded, while an overall agreement with the $\Lambda$CDM predictions on $z_{tr}$ still persists.

Summing up, in both the two methods, we infer the main physical conclusions reported below.

\begin{itemize}
\item[{\bf A.}] The most feasible approximations suggest transition redshift compatible intervals than the $\Lambda$CDM predictions, though our findings cannot exclude dark energy at $2\sigma$ confidence level.
\item[{\bf B.}] The multiplicative and additive degeneracy among coefficients is mainly responsible for the overall error bars estimated in our fits, while using the OHD catalog favors a hierarchy up to $j_{tr}$.
\item[{\bf C.}] Error bars are not significantly underestimated and  at $1\sigma$ we find a slight concordance with the standard cosmological model for $H_0$ and $z_{tr}$. On the other hand, at $2\sigma$ confidence levels the trend of compatibility with the standard model is no longer valid.
\item[{\bf D.}] The cosmographic bounds on the jerk and snap coefficients strongly disagree, even within the $1\sigma$, with the standard cosmological prescriptions at the transition time.
\item[{\bf E.}] Evaluating the regions in which $z_{eq}$ spans, within the expectations over $z_{tr}$, leads to exclusions plots that are compatible with the $\Lambda$CDM predictions, but do not exclude, as well as for the transition redshift, a dark energy evolution, different from a pure cosmological constant.
\end{itemize}

From the above summary, we define the most suitable values for $H_0, z_{tr}, j_{tr}$ and $s_{tr}$ below, taking the averages of each hierarchy. The first results below are for hierarchy 1, {\it i.e.}, lowest $\mathcal N$.

\begin{subequations}
\begin{align}
&{\rm {\bf DHE\,\,\, method}} \nonumber\\
&{\rm \,\,\,}\nonumber\\
&{\rm SN+OHD+BAO} \nonumber\\
&\bar z_{tr}=0.651^{+0.025}_{-0.158}\,,\\
&{\rm OHD}\nonumber\\
&\bar z_{tr}=0.566^{+0.371}_{-0.124}\,,\\
&{\rm \,\,\,}\nonumber\\
&{\rm {\bf DDPE\,\,\, method}}\nonumber\\
&{\rm \,\,\,}\nonumber\\
&{\rm SN+OHD+BAO} \nonumber\\
&\bar z_{tr}=1.0215^{+0.013}_{-0.173}\,,\\
&{\rm OHD}\nonumber\\
&\bar z_{tr}=0.595^{+0.260}_{-0.119}\,,
\end{align}
\end{subequations}

\noindent and for hierarchy 2, {\it i.e.}, largest $\mathcal N$, we have

\begin{subequations}
\begin{align}
&{\rm {\bf DHE\,\,\, method}}\nonumber\\
&{\rm \,\,\,}\nonumber\\
&{\rm SN+OHD+BAO} \nonumber\\
&\bar z_{eq}\in[0.211; 0.406]\,,\\
&{\rm OHD}\nonumber\\
&\bar z_{eq}\in[0.189; 0.562]\,,\\
&{\rm \,\,\,}\nonumber\\
&{\rm {\bf DDPE\,\,\, method}} \nonumber\\
&{\rm \,\,\,}\nonumber\\
&{\rm SN+OHD+BAO} \nonumber\\
&\bar z_{eq}\in[0.364; 0.621]\,,\\
&{\rm OHD}\nonumber\\
&\bar z_{eq}\in [0.204; 0.513]\,.
\end{align}
\end{subequations}

Clearly our approach does not fully exclude the CPL parametrization and the $\omega$CDM paradigm. Moreover the hypothesis of a dark fluid is also plausible as clearly shown above. Such models fully degenerate with respect to the concordance paradigm, but we believe, in view of the {\it Occam razor}, the $\Lambda$CDM would be the statistically favored one. For the sake of completeness, however, more investigations with additional data sets would be essential to heal the degeneracy among models and to fully conclude which model is effectively the best framework predicted by the set of transition and equivalence redshifts.



\section{Conclusions}\label{sez7}
In this work, we investigated two possible model independent methods to estimate the transition and equivalence redshifts. The first redshift is associated with the time at which the deceleration parameter changes sign, whereas the second is associated with the equivalence of the matter and dark energy densities. The strategy was based on cosmographic Taylor expansions of the Hubble rate and deceleration parameters respectively. In particular, we first expanded the Hubble rate around $z_{tr}$ up to two hierarchies and then we worked out the same for the deceleration parameter $q(z)$, again with two hierarchies. Both the expansions have been evaluated up to the snap parameter at the transition.

The corresponding fits have been performed using $z_{tr}$ expansions only by virtue of the requirement $q(z_{tr})=0$, that mostly simplified our computation. We thus involved two types of fits based on OHD data first and then combining SNe Ia with OHD and BAO. We then computed the luminosity distance, the angular distance and the Hubble rate expanded with the aforementioned hierarchies. We evaluated a set of constraints taking free all the coefficients, and adopting MCMC simulations, making use of the Metropolis-Hastings algorithm.

Even though this technique led to promising results, we obtained large systematics that have been discussed throughout the text. Even if we did not underestimate the error bars, the concordance paradigm is inside our results within $1\sigma$, but deviations seemed to be more evident at $2\sigma$ confidence level. However, even if the corresponding $\Lambda$CDM predictions looked plausible at $1\sigma$ confidence level only, we believe our treatments could be strongly plagued by the approximation we made and by fixing the hierarchy $\mathcal N$. Consequently, the concordance paradigm results were not excluded, being in a good agreement at least within the limits we imposed. For the sake of completeness, our approach did not fully exclude that dark energy, within the redshift $z<1$, could slightly evolve. This fact could be subject of refined analyses that will be carried forward, increasing the fit accuracy or simply adopting a high-redshift cosmographic analysis. To stress how much the concordance paradigm matched our results, we also computed exclusion regions in which $z_{eq}$ is not allowed to span. We showed such regions are again compatible with the standard paradigm, but still not completely able to exclude an evolving dark energy term that degenerates with the $\Lambda$CDM paradigm. Finally, we noticed that fixing suitable intervals of $(z_{tr}, z_{eq})$ implies we can use them as \emph{discriminators for dark energy models}. In other words, we can improve our analyses, fixing tighter intervals over $(z_{tr}, z_{eq})$, and investigating \emph{de facto} whether dark energy is in the form of a pure cosmological constant or  weakly evolve.

Thus, future fits are expected to improve the quality of current ones, by adding further orders beyond what is presented here. We therefore expect to improve our treatment by including additional catalogs of data and to confront specific dark energy models with the DHE and DDPE methods.

\section*{Acknowledgements}
SC acknowledges the contribution of INFN (IS-QGSKY, IS-MOONLight2) for financial support. PKSD acknowledges support from the First Rand Bank, South
Africa. OL expresses his gratitude to Dr. Marco Muccino and to the Ministry of Education and Science of the Republic of Kazakhstan,  Grant: IRN AP08052311 that partially supported this work.

\section*{Data availability}
SN and BAO data are available in \citet{ML} and \citet{z0} respectively whereas OHD data are available in \citet{datiohd}.

\bibliographystyle{mnras}

\newpage

\appendix


\newpage

 {\centerline{\bf Appendix A:   COSMOGRAPHIC SERIES}}\label{appendiceA}\vspace{0.3cm}

\noindent The luminosity distance is expanded around $z\simeq0$ as

\begin{subequations}\label{dlespansa}
\begin{align}
    d_L^{(3)}(q_0,j_0,z) & =  \frac{1}{H_0} \Bigl[ z + z^2  \Bigl(\frac{1}{2} - \frac{q_0}{2} \Bigr) +
    z^3  \Bigl(-\frac{1}{6} -\frac{j_0}{6} + \frac{q_0}{6} + \frac{q_0^2}{2} \Bigr)\Bigr]\,,\\
    d_L^{(4)}(q_0,j_0,z) & =  d_L(q_0,j_0,z)^{(3)} +\\
    &z^4  \Bigl( \frac{1}{12} + \frac{5 j_0}{24} - \frac{q_0}{12} + \frac{5 j_0 q_0}{12} -
    \frac{5 q_0^2}{8} - \frac{5 q_0^3}{8} + \frac{s_0}{24} \Bigr)\,.
\end{align}
\end{subequations}

\noindent The Hubble rate is expanded around $z\simeq0$ by

\begin{subequations}\label{Hespanso}
\begin{align}
    H^{(2)}(q_0,j_0,z) & = H_0\left[1+(1+q_0)z+\frac{z^2}{2}(j_0-q_0^2)\right] \,,\\
    H^{(3)}(q_0,j_0,s_0,z) & = H^{(2)}(q_0,j_0,z) + z^3\left(3 q_0^2 + 3 q_0^3 - j_0 (3 + 4 q_0) - s_0 \right)  \,.
\end{align}
\end{subequations}

\noindent The angular distance expansions around $z\simeq0$ read

\begin{subequations}\label{dAespansa}
\begin{align}
    d_A^{(2)}(q_0,j_0,z) & =
    100 r_s \frac{h_0}{cz}\times\nonumber\\
    &\times\left[1 + \frac{2}{3}(1 + q_0) z + \frac{1}{36}(-3 + 10 j_0 - 6 q_0 - 13 q_0^2) z^2\right]\,,\\
    &d_A^{(3)}(q_0,j_0,s_0,z)  =  d_L^{(2)}(q_0,j_0,z) +\\ &\frac{z^3}{324}(13 - 111 j_0 + 39 q_0 - 138 j_0 q_0 + 150 q_0^2 + 124 q_0^3 - 27 s_0)\nonumber\,.
\end{align}
\end{subequations}

 {\centerline{\bf Appendix B:  BAO AND OHD DATA POINTS}}\vspace{0.3cm}

In this appendix, we list the data points used in the paper, besides SNe Ia. In particular, below we report the BAO and OHD catalogs.

\begin{table}
\setlength{\tabcolsep}{0.3em}
\renewcommand{\arraystretch}{1.0}
\begin{tabular}{l c c l}
\hline\hline
Survey & $z$ & $d_{\rm z}$ & Ref.\\
\hline
6dFGS & $0.106$ & $0.3360\pm0.0150$ & \cite{2011MNRAS.416.3017B}\\
SDSS-DR7 & $0.15$ & $0.2239\pm0.0084$ & \cite{2015MNRAS.449..835R}\\
SDSS &  $0.20$ & $0.1905\pm0.0061$ & \cite{2010MNRAS.401.2148P}\\
SDSS-III & $0.32$ & $0.1181\pm0.0023$ & \cite{2014MNRAS.441...24A}\\
SDSS & $0.35$ & $0.1097\pm0.0036$ & \cite{2010MNRAS.401.2148P}\\
SDSS-III & $0.57$ & $0.0726\pm0.0007$ & \cite{2014MNRAS.441...24A}\\
SDSS-III & $2.34$ & $0.0320\pm0.0016$ & \cite{2015AA...574A..59D}\\
SDSS-III & $2.36$ & $0.0329\pm0.0012$ & \cite{2014JCAP...05..027F}\\
\hline
\hline
\end{tabular}
\caption{Numerical values take from uncorrelated BAO points with the corresponding references.}
\label{tab:BAO}
\end{table}

\begin{table}
\setlength{\tabcolsep}{1.5em}
\small
\caption{$H(z)$ measurements got from differential age treatment, in which $H(z)$ and error bars, $\sigma_H$, are given in units km/s/Mpc.}
{\begin{tabular}{c c }
\hline
\hline
 $z$ &$H \pm \sigma_H$  \\
\hline
0.0708	& $69.00 \pm 19.68$  \\
0.09	& $69.0 \pm 12.0$  \\
0.12	& $68.6 \pm 26.2$  \\
0.17	& $83.0 \pm 8.0$  \\
0.179 & $75.0 \pm	4.0$  \\
0.199 & $75.0	\pm 5.0$  \\
0.20 &$72.9 \pm 29.6$  \\
0.27	& $77.0 \pm 14.0$  \\
0.28	& $88.8 \pm 36.6$  \\
0.35	& $82.1 \pm 4.85$ \\
0.352 & $83.0	\pm 14.0$  \\
0.3802	& $83.0 \pm 13.5$ \\
0.4 & $95.0	\pm 17.0$  \\
0.4004	& $77.0 \pm 10.2$  \\
0.4247	& $87.1 \pm 11.2$   \\
0.4497 &	$92.8 \pm 12.9$ \\
0.4783	 & $80.9 \pm 9.0$  \\
0.48	& $97.0 \pm 62.0$  \\
0.593 & $104.0 \pm 13.0$  \\
0.68	& $92.0 \pm 8.0$  \\
0.781 & $105.0 \pm 12.0$  \\
0.875 & $125.0 \pm 17.0 $  \\
0.88	& $90.0 \pm 40.0$  \\
0.9 & $117.0 \pm 23.0$   \\
1.037 & $154.0 \pm 20.0$  \\
1.3 & $168.0 \pm 17.0$  \\
1.363 & $160.0 \pm 33.6$  \\
1.43	& $177.0 \pm18.0$  \\
1.53	& $140.0	\pm 14.0$  \\
1.75	 & $202.0 \pm 40.0$  \\
1.965& $186.5 \pm 50.4$  \\
\hline
\hline
\end{tabular}
 \label{tab:OHD}
 }
\end{table}

\newpage

{\centerline{\bf Appendix C:   CONTOUR PLOTS}}\vspace{0.3cm}

\noindent In this section, we report the contours got from our analyses.

\subsection*{DHE method}

\begin{figure*}
\centering
\includegraphics[height=0.25\hsize,clip]{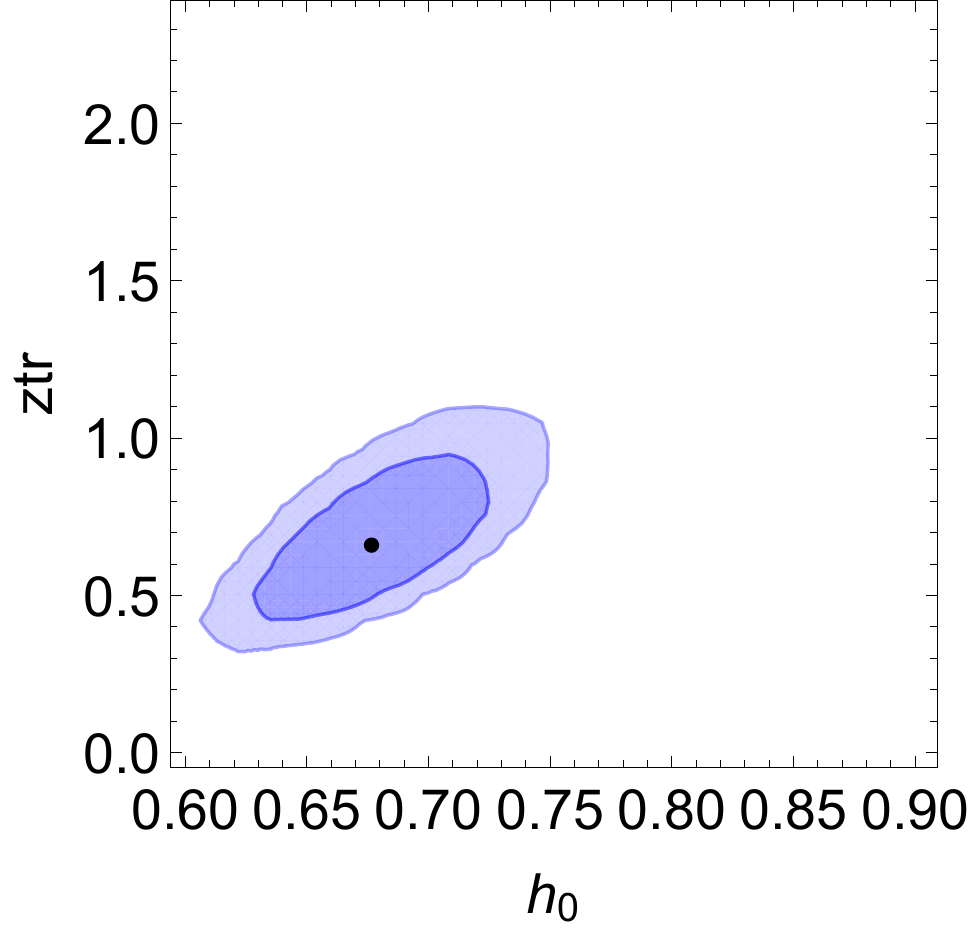}
\includegraphics[height=0.25\hsize,clip]{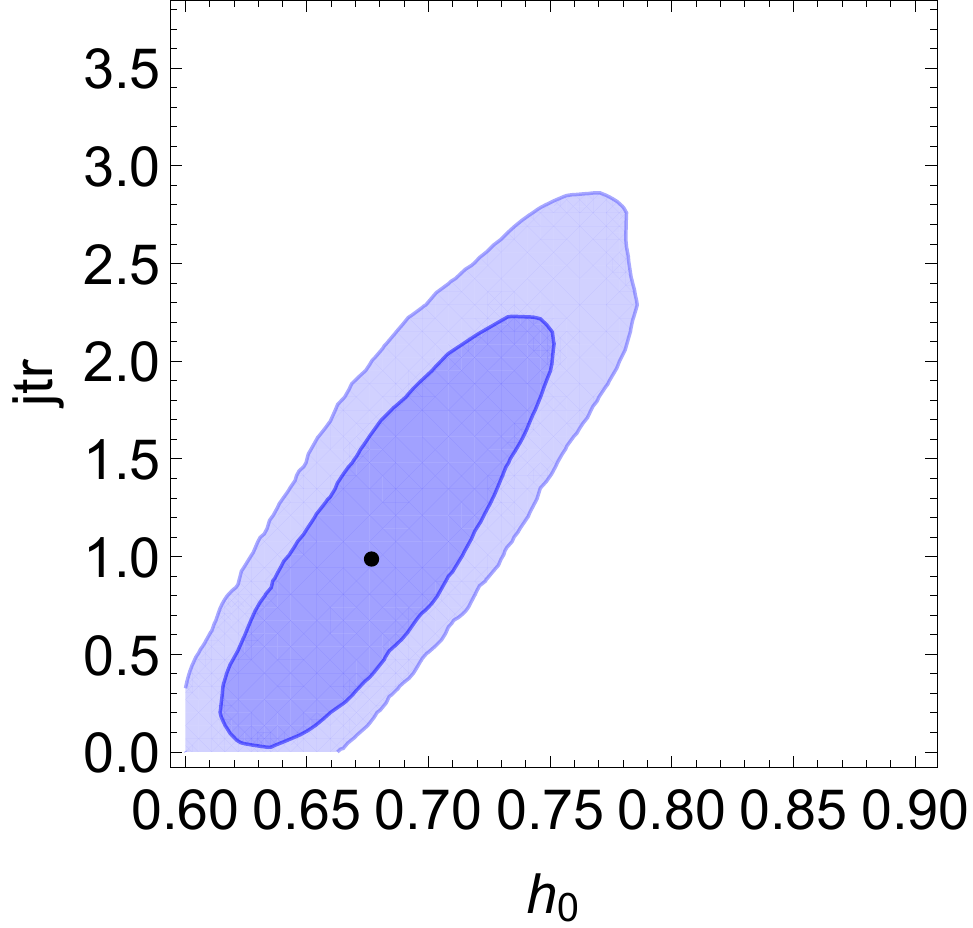}
\includegraphics[height=0.25\hsize,clip]{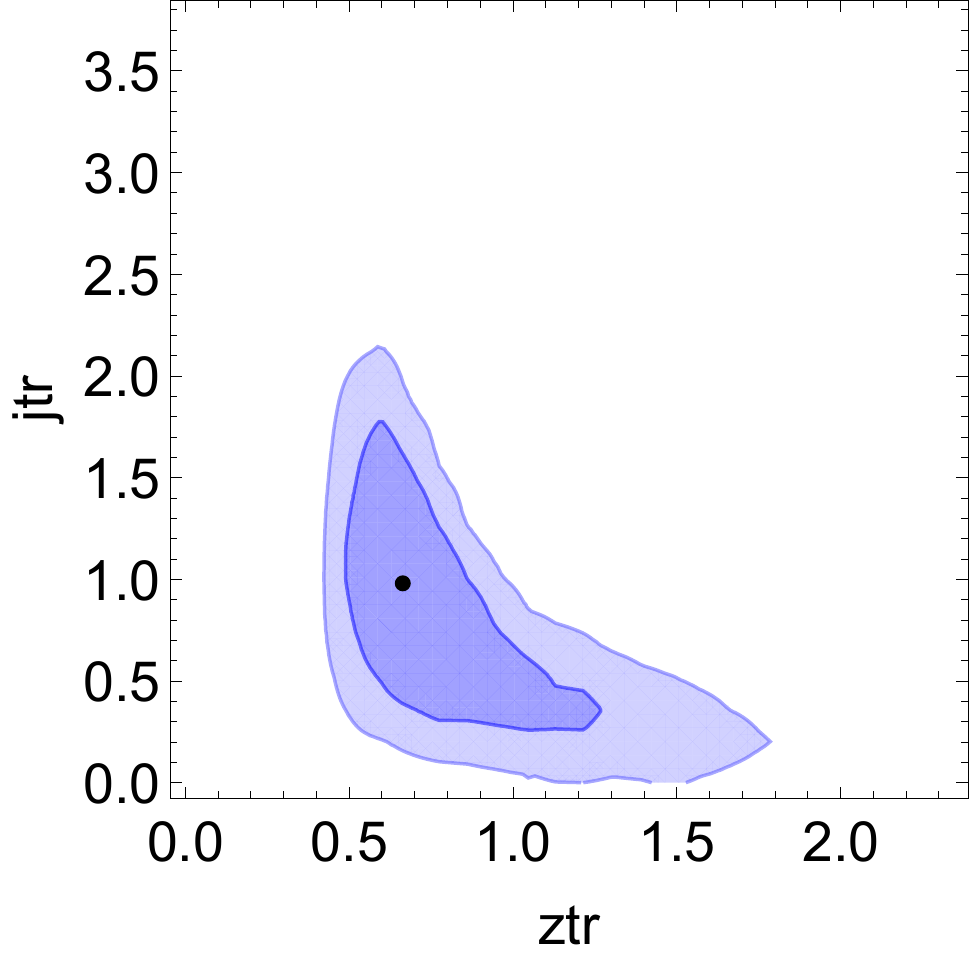}
\caption{Contours for the DHE method using the OHD data set and the second order expansion of the Hubble parameter. }
\label{contourDHEh1OHD}
\end{figure*}

\begin{figure*}
\centering
\includegraphics[height=0.25\hsize,clip]{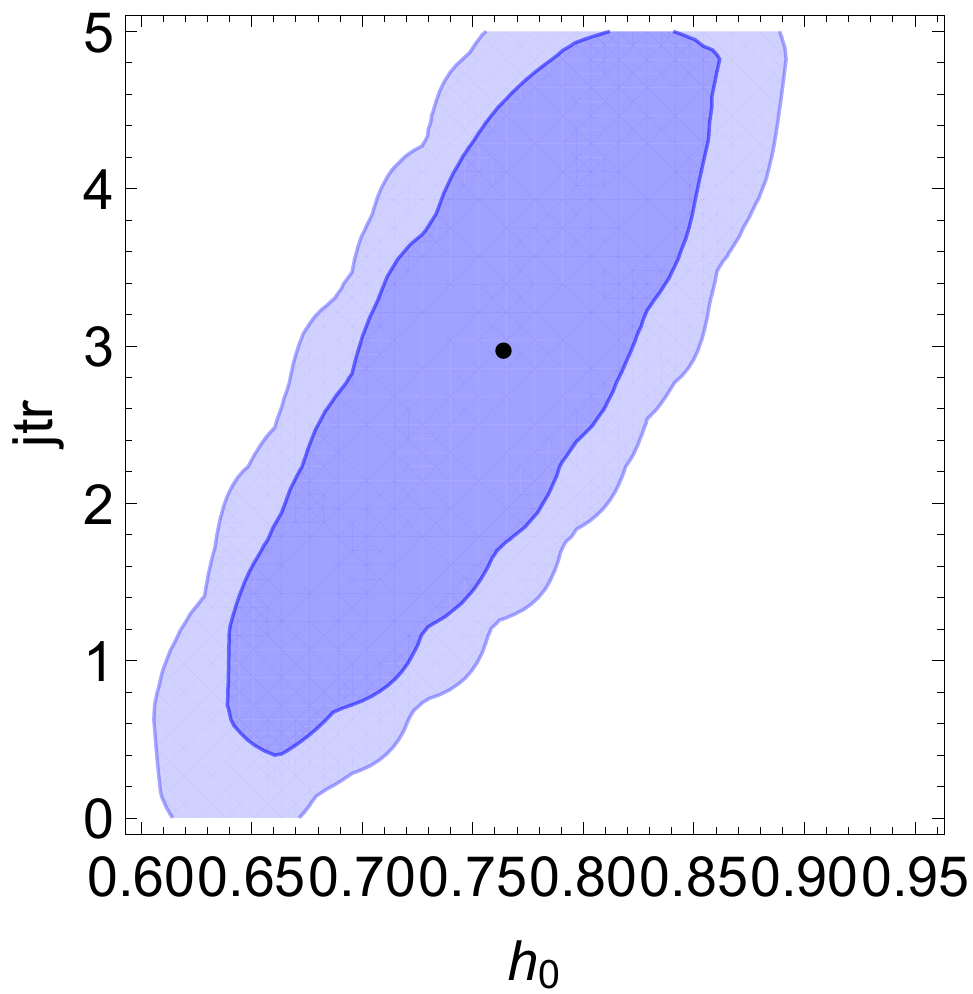}
\includegraphics[height=0.25\hsize,clip]{DHE_h3_OHD_jtrVSh0.pdf}
\includegraphics[height=0.25\hsize,clip]{DHE_h3_OHD_jtrVSh0.pdf}
\includegraphics[height=0.25\hsize,clip]{DHE_h3_OHD_jtrVSh0.pdf}
\includegraphics[height=0.25\hsize,clip]{DHE_h3_OHD_jtrVSh0.pdf}
\includegraphics[height=0.25\hsize,clip]{DHE_h3_OHD_jtrVSh0.pdf}
\caption{Contours for the DHE method using the OHD data set and the third order expansion of the Hubble parameter. }
\label{contourDHEh2OHD}
\end{figure*}

\noindent The contours using the OHD data are got in Figs. \ref{contourDHEh1OHD} and \ref{contourDHEh2OHD}.

\begin{figure*}
\centering
\includegraphics[height=0.25\hsize,clip]{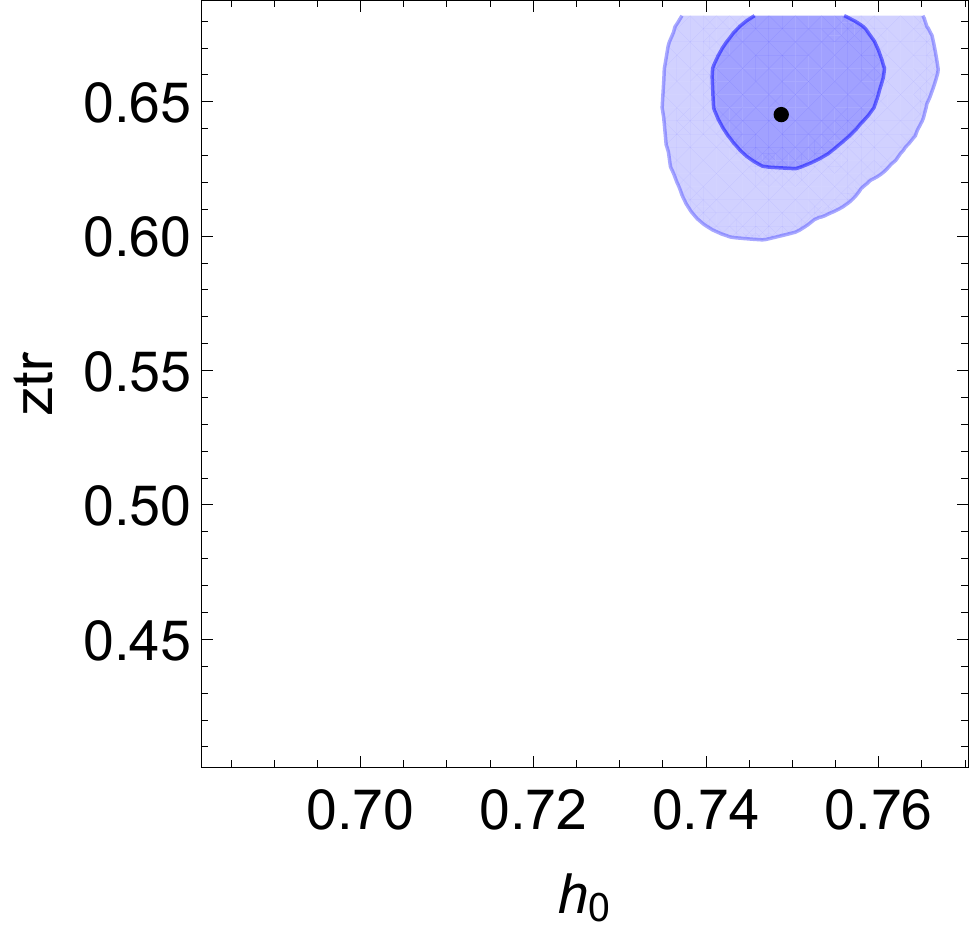}
\includegraphics[height=0.25\hsize,clip]{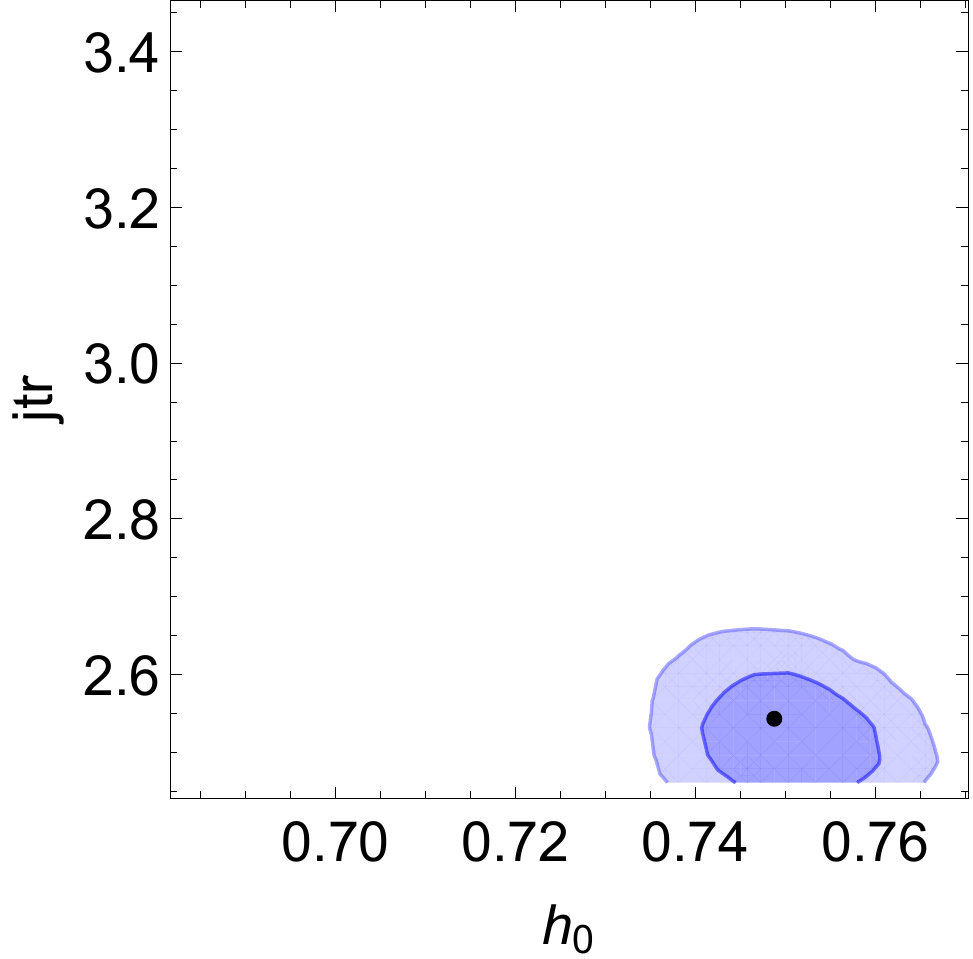}
\includegraphics[height=0.25\hsize,clip]{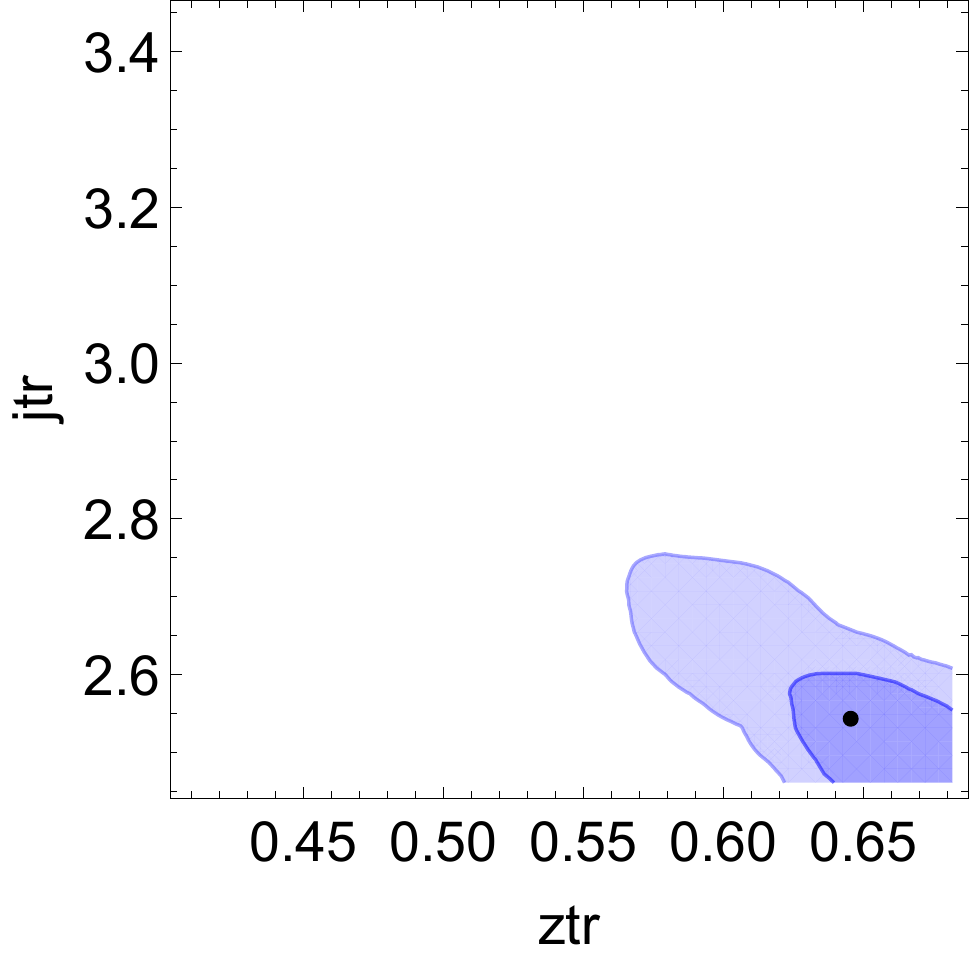}
\caption{Contours for the DHE method using the SN+OHD+BAO data sets and the second order expansion of the Hubble parameter. }
\label{contourDHEh2SNOHDBAO}
\end{figure*}

\begin{figure*}
\centering
\includegraphics[height=0.25\hsize,clip]{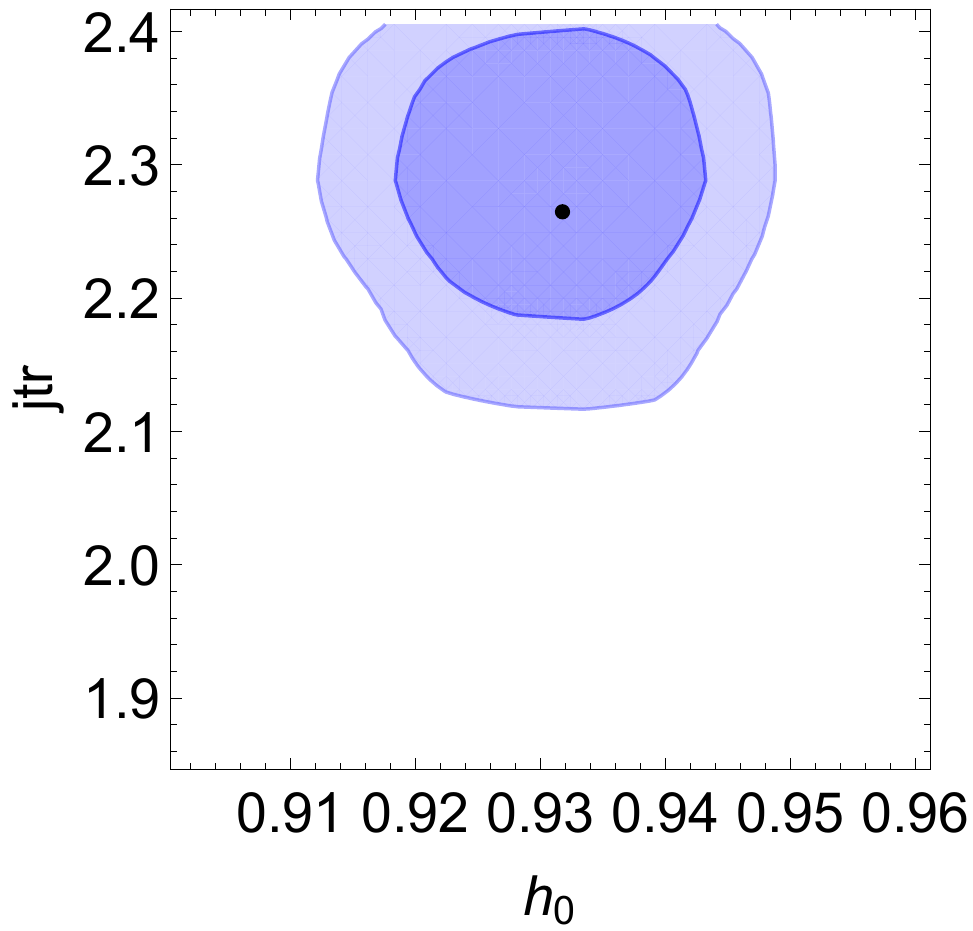}
\includegraphics[height=0.25\hsize,clip]{DHE_h3_SNOHDBAO_jtrVSh0.pdf}
\includegraphics[height=0.25\hsize,clip]{DHE_h3_SNOHDBAO_jtrVSh0.pdf}
\includegraphics[height=0.25\hsize,clip]{DHE_h3_SNOHDBAO_jtrVSh0.pdf}
\includegraphics[height=0.25\hsize,clip]{DHE_h3_SNOHDBAO_jtrVSh0.pdf}
\includegraphics[height=0.25\hsize,clip]{DHE_h3_SNOHDBAO_jtrVSh0.pdf}
\caption{Contours for the DHE method using the SN+OHD+BAO data sets and the third order expansion of the Hubble parameter. }
\label{contourDHEh3SNOHDBAO}
\end{figure*}

\noindent The contours using SNe Ia with OHD and BAO catalogs are portrayed in Figs. \ref{contourDHEh2SNOHDBAO} and \ref{contourDHEh3SNOHDBAO}.

\subsection*{DDPE method}

\begin{figure*}
\centering
\includegraphics[height=0.25\hsize,clip]{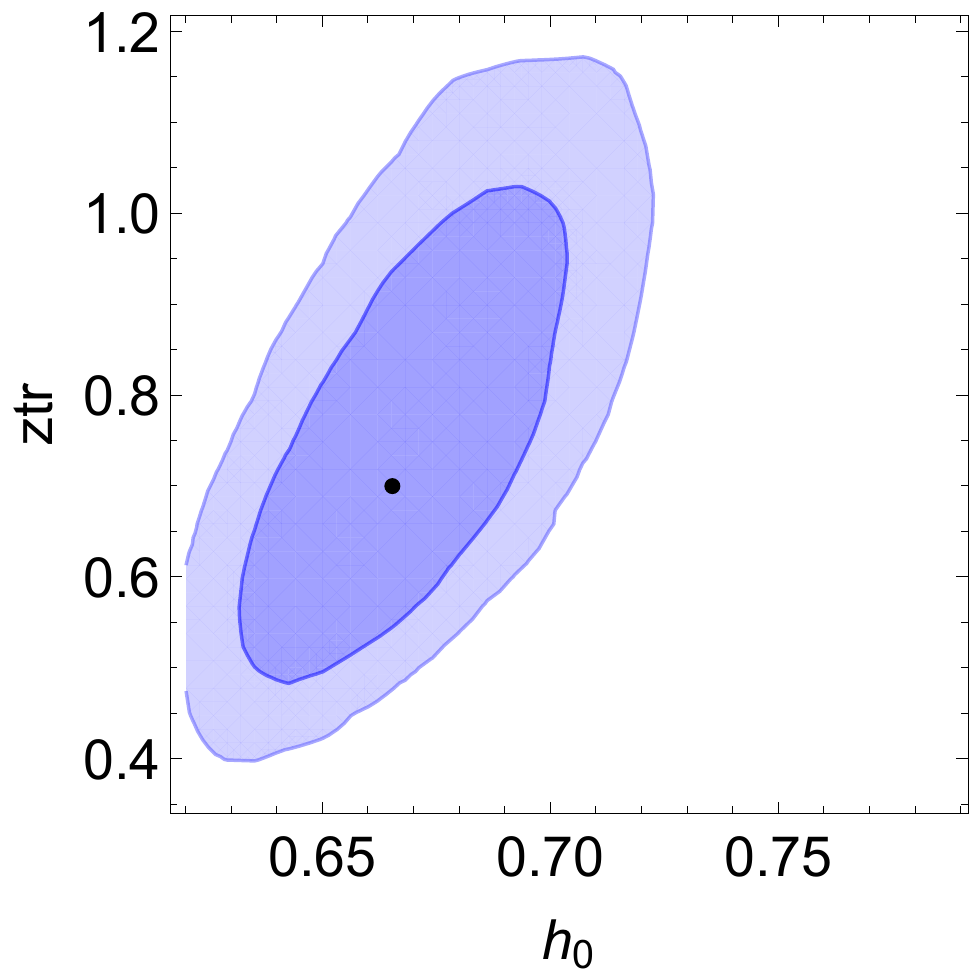}
\includegraphics[height=0.25\hsize,clip]{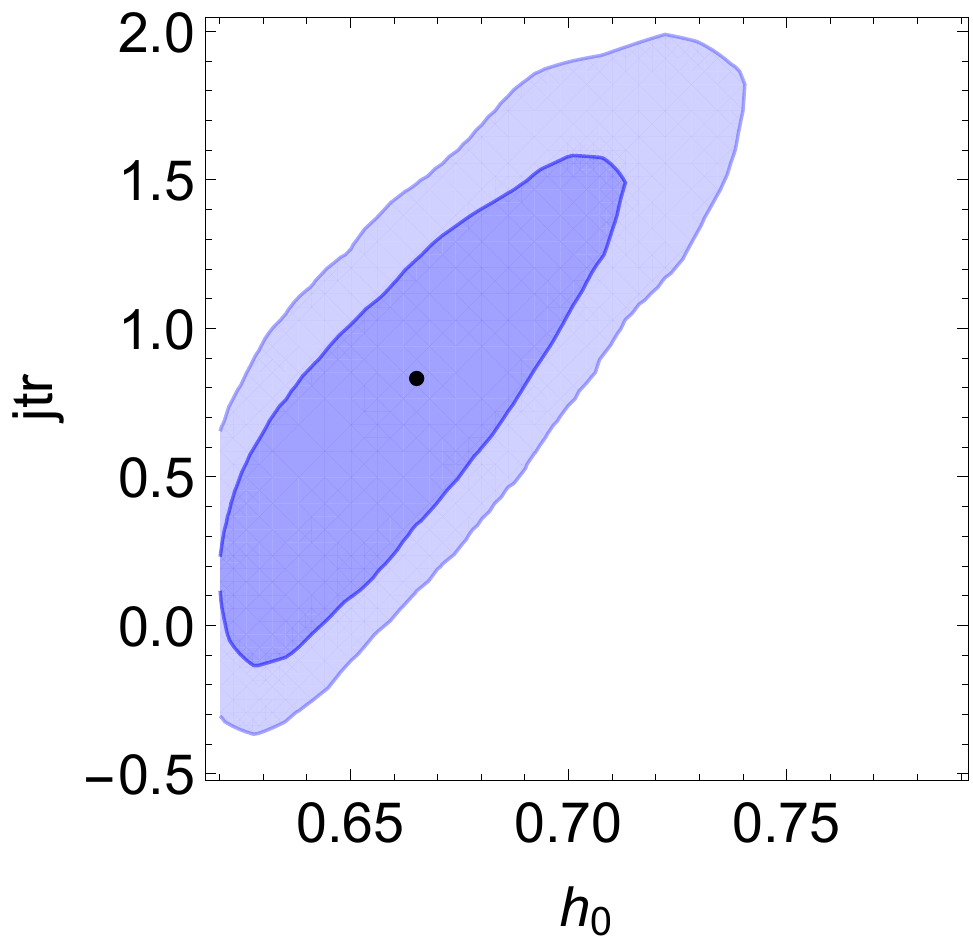}
\includegraphics[height=0.25\hsize,clip]{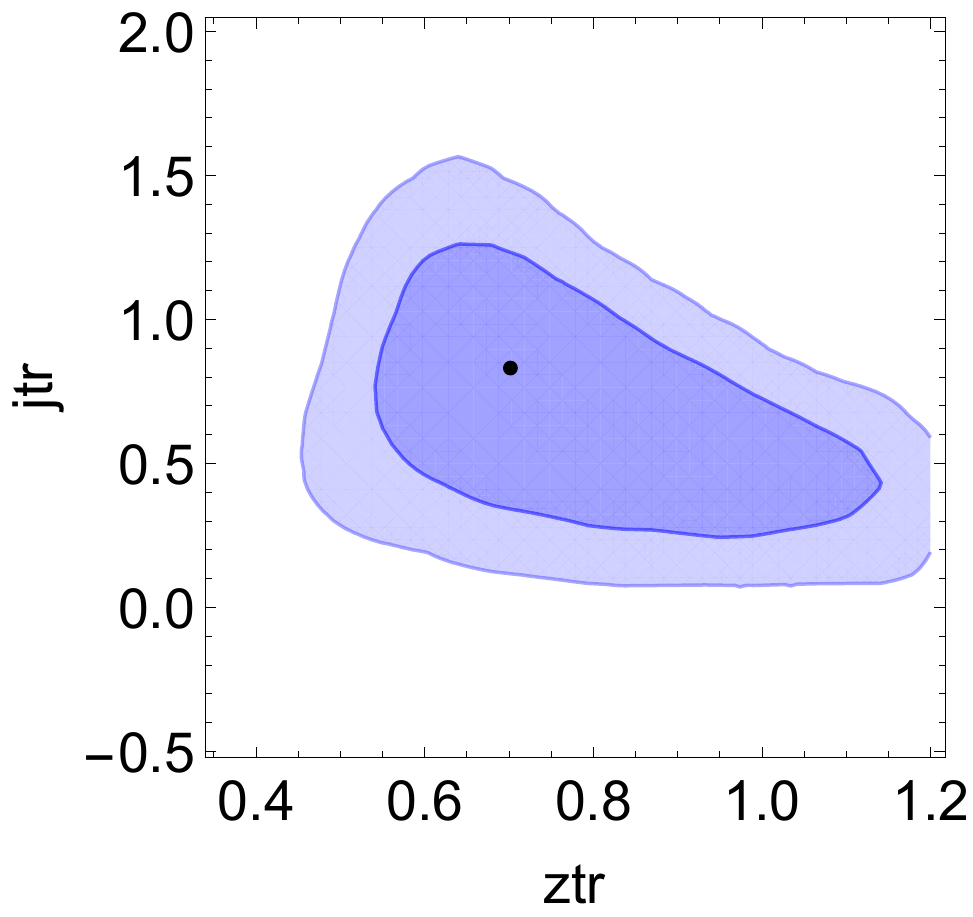}
\caption{Contours for the DDPE method using the OHD data set and the first order expansion of the deceleration parameter. }
\label{contourDDPEq1OHD}
\end{figure*}

\begin{figure*}
\centering
\includegraphics[height=0.25\hsize,clip]{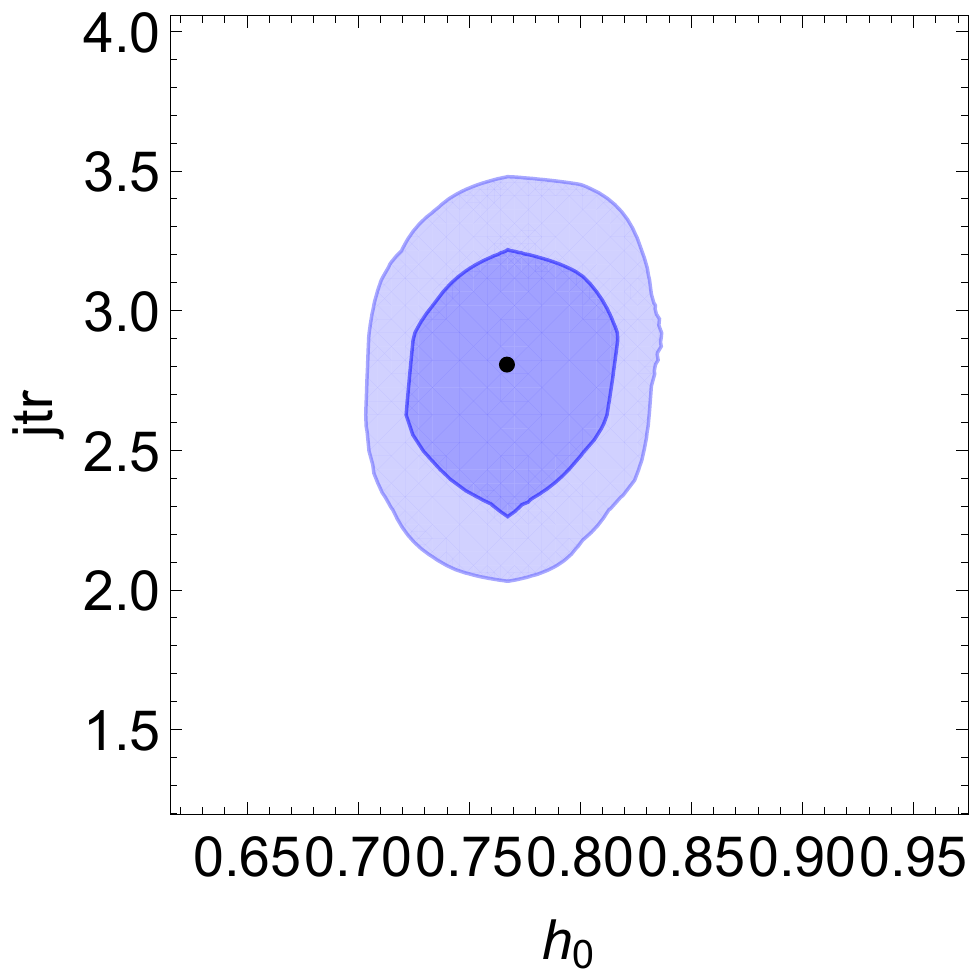}
\includegraphics[height=0.25\hsize,clip]{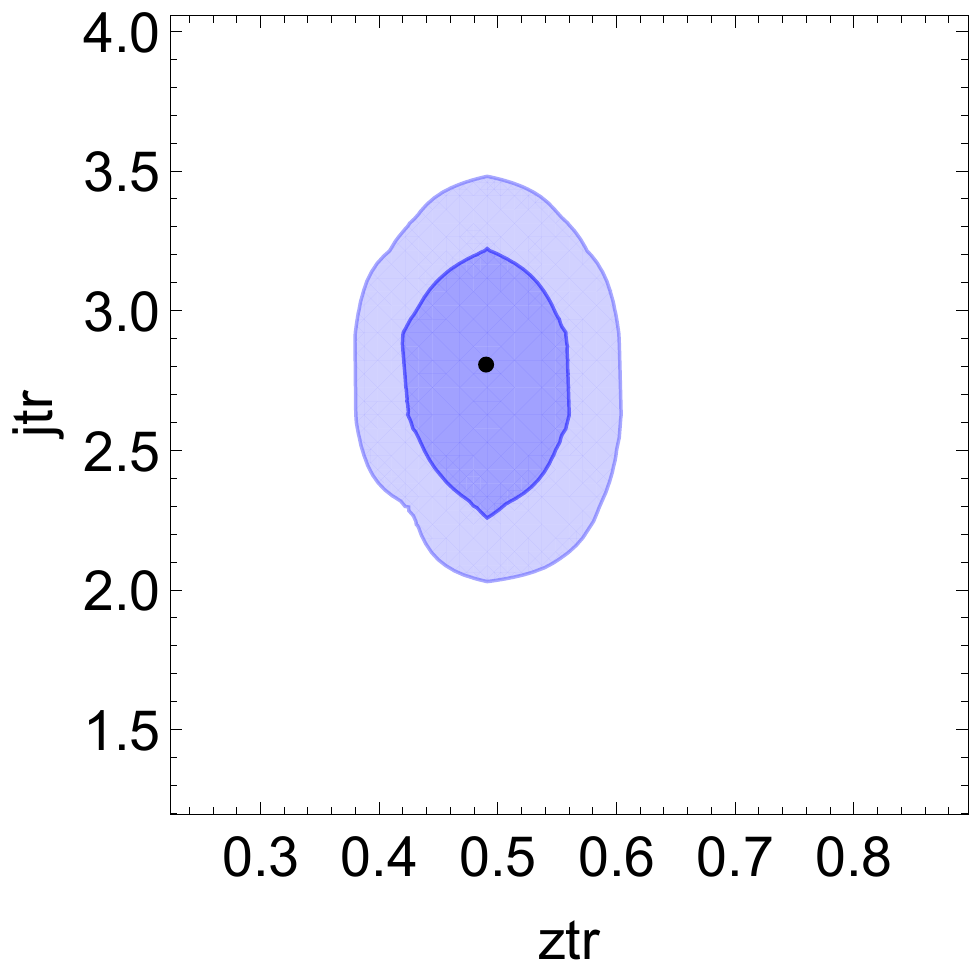}
\includegraphics[height=0.25\hsize,clip]{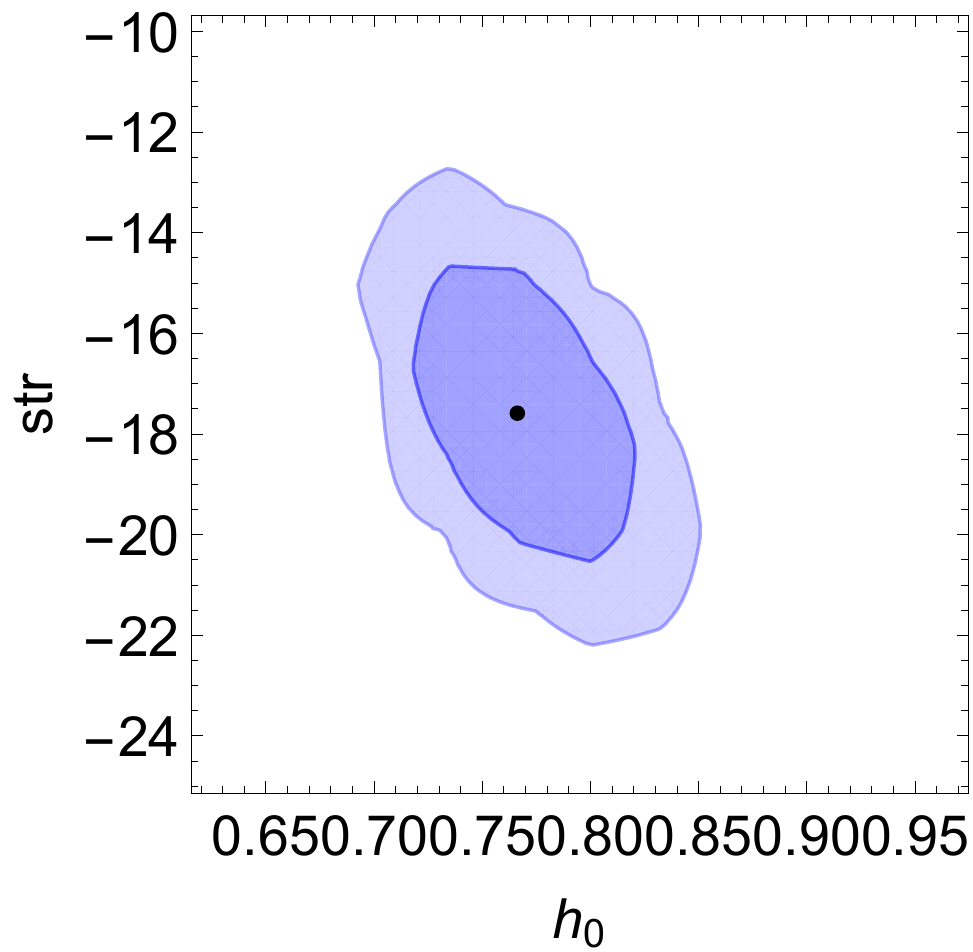}
\includegraphics[height=0.25\hsize,clip]{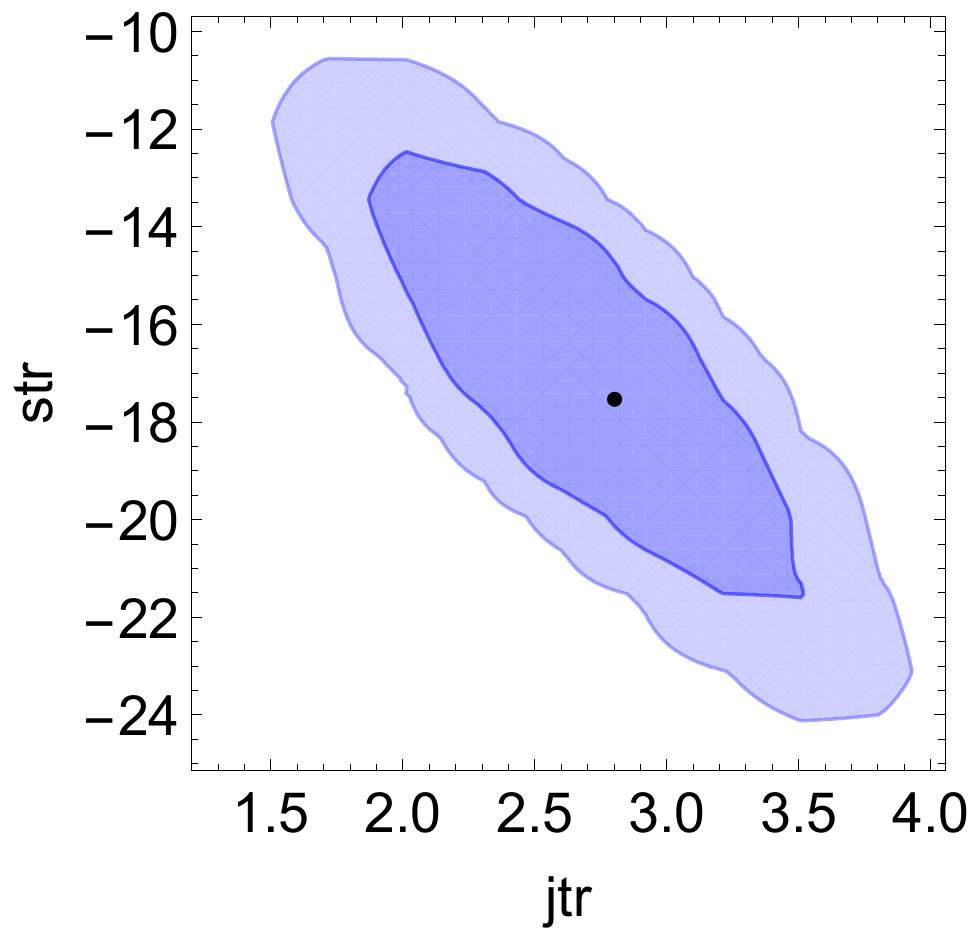}
\includegraphics[height=0.25\hsize,clip]{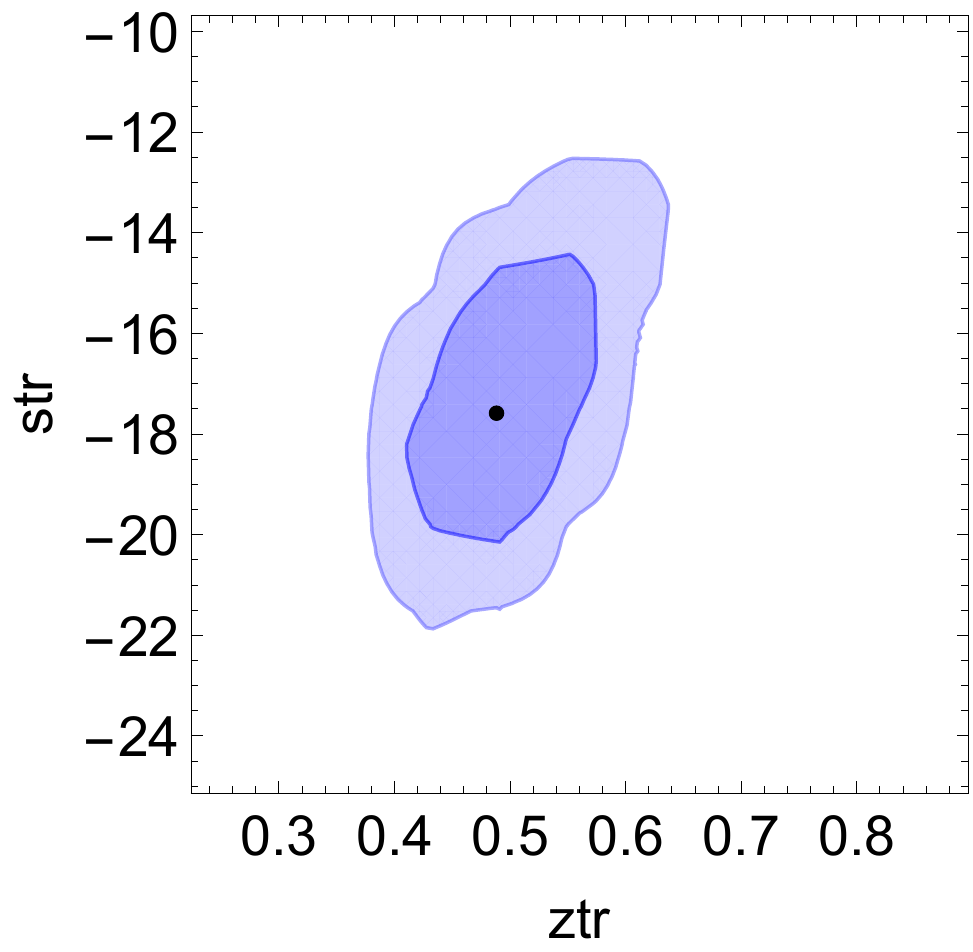}
\includegraphics[height=0.25\hsize,clip]{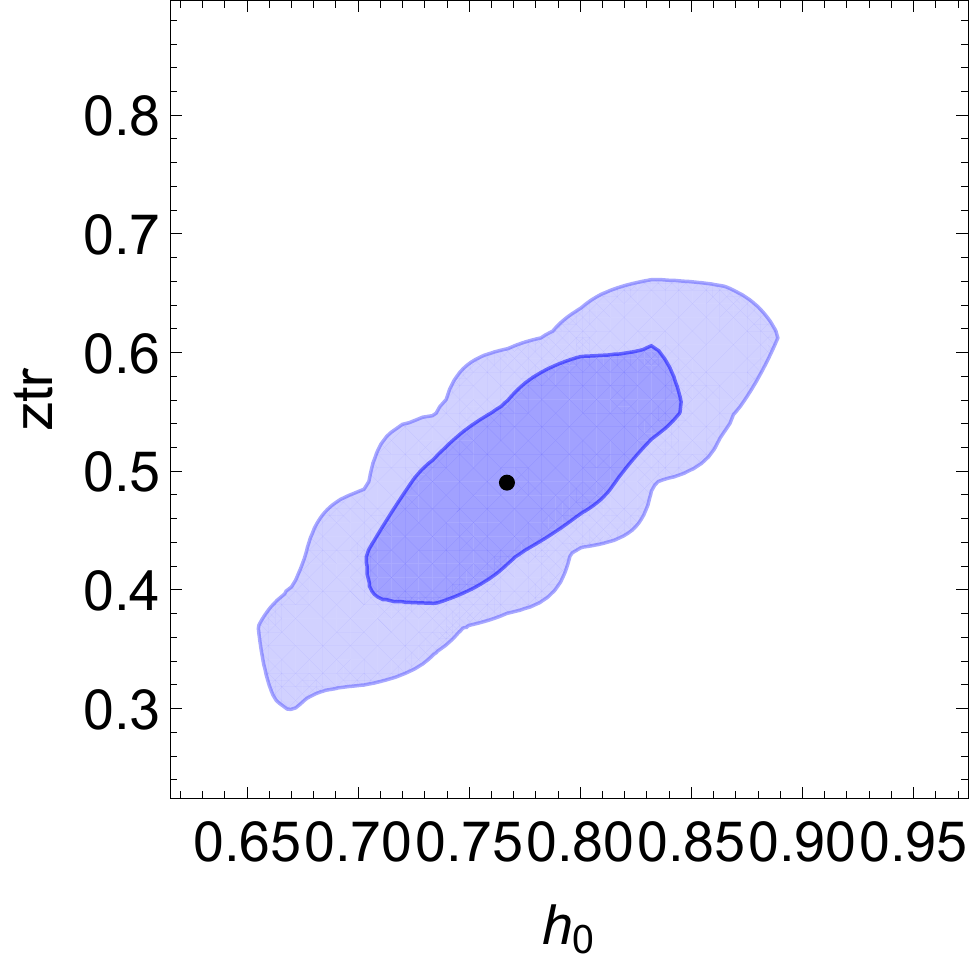}
\caption{Contours for the DDPE method using the OHD data set and the second order expansion of the deceleration parameter. }
\label{contourDDPEq2OHD}
\end{figure*}

\noindent The contours evaluated using OHD are displayed in Figs. \ref{contourDDPEq1OHD} and \ref{contourDDPEq2OHD}.

\begin{figure*}
\centering
\includegraphics[height=0.25\hsize,clip]{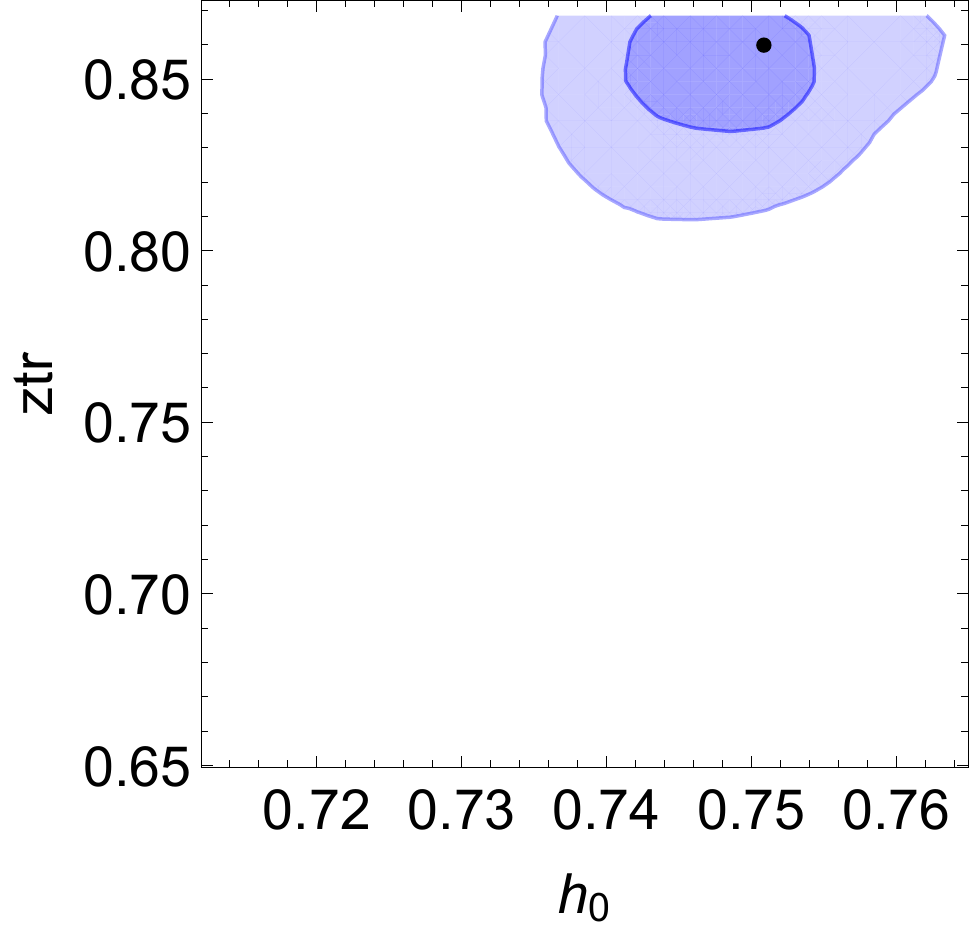}
\includegraphics[height=0.25\hsize,clip]{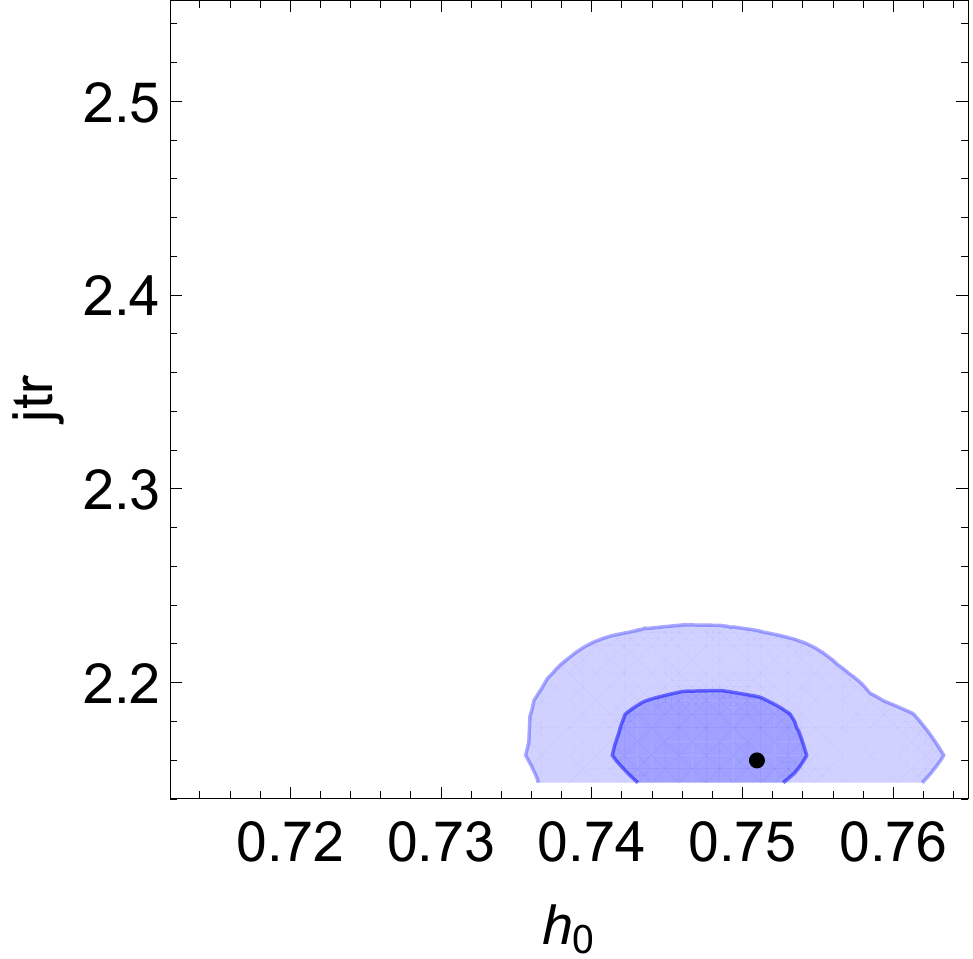}
\includegraphics[height=0.25\hsize,clip]{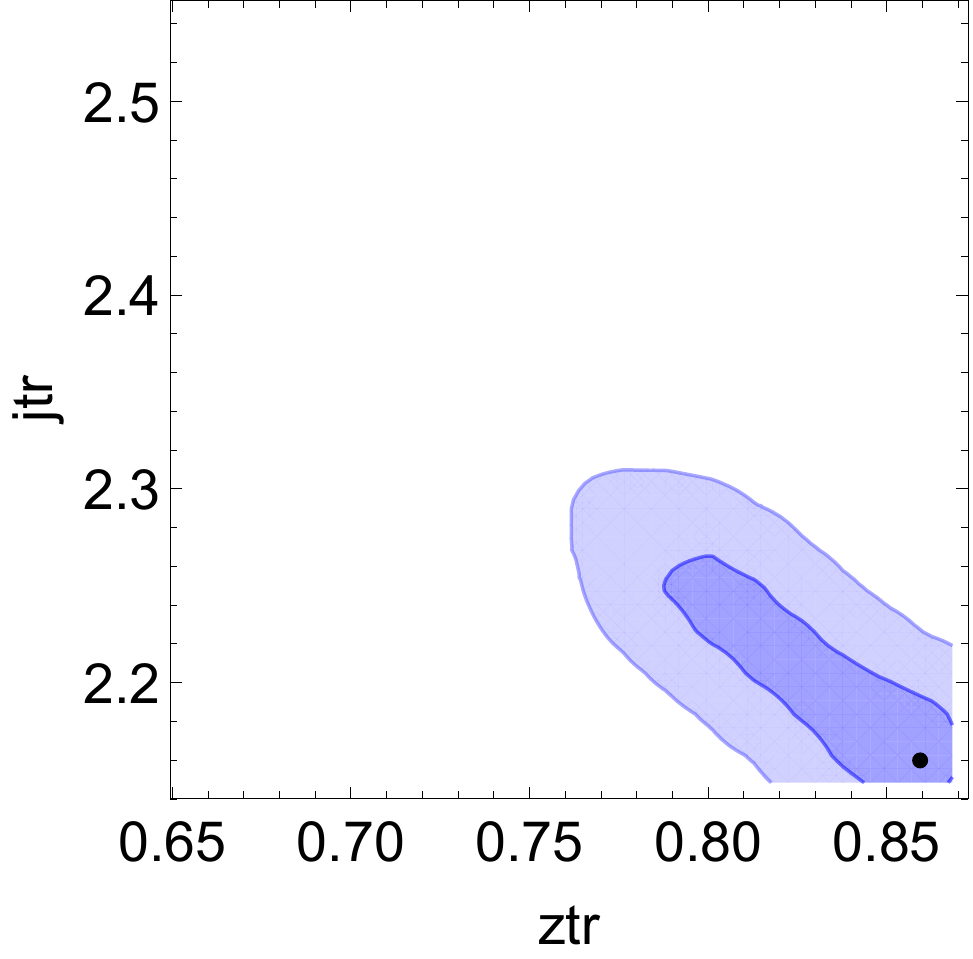}
\caption{Contours for the DDPE method using the SN+OHD+BAO data sets and the first order expansion of the deceleration parameter. }
\label{contourDDPEq1SNOHDBAO}
\end{figure*}

\begin{figure*}
\centering
\includegraphics[height=0.25\hsize,clip]{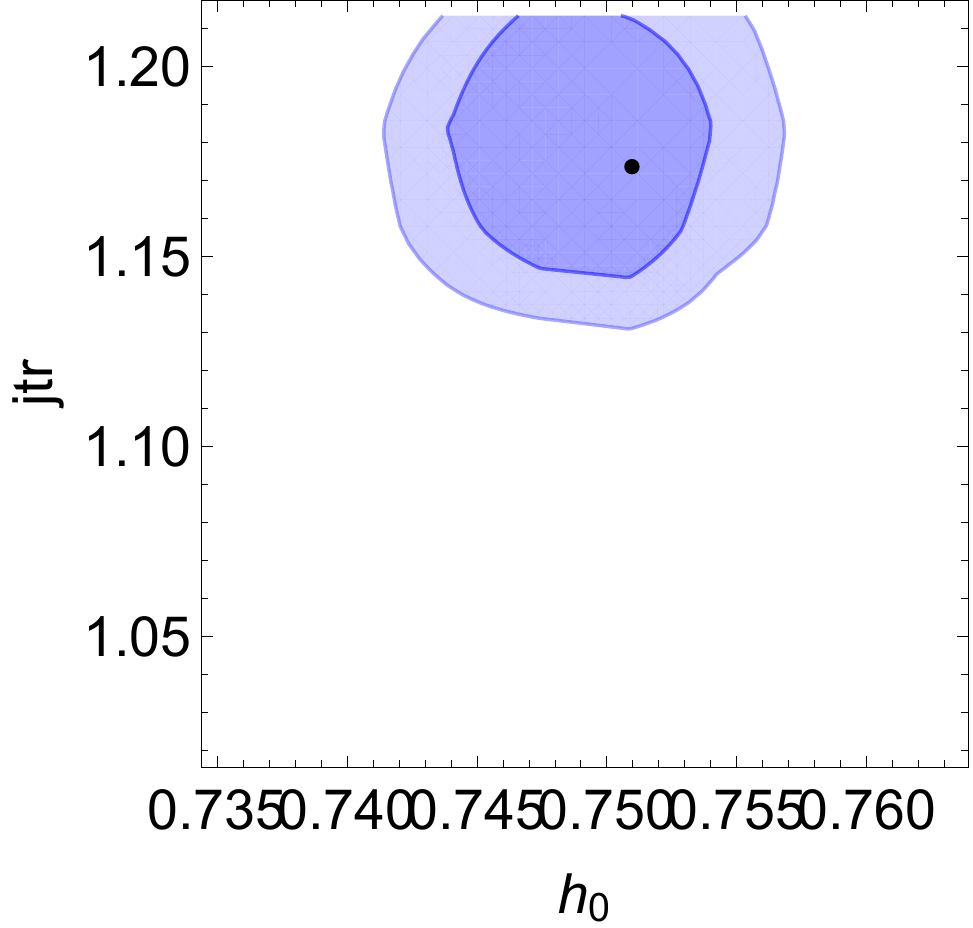}
\includegraphics[height=0.25\hsize,clip]{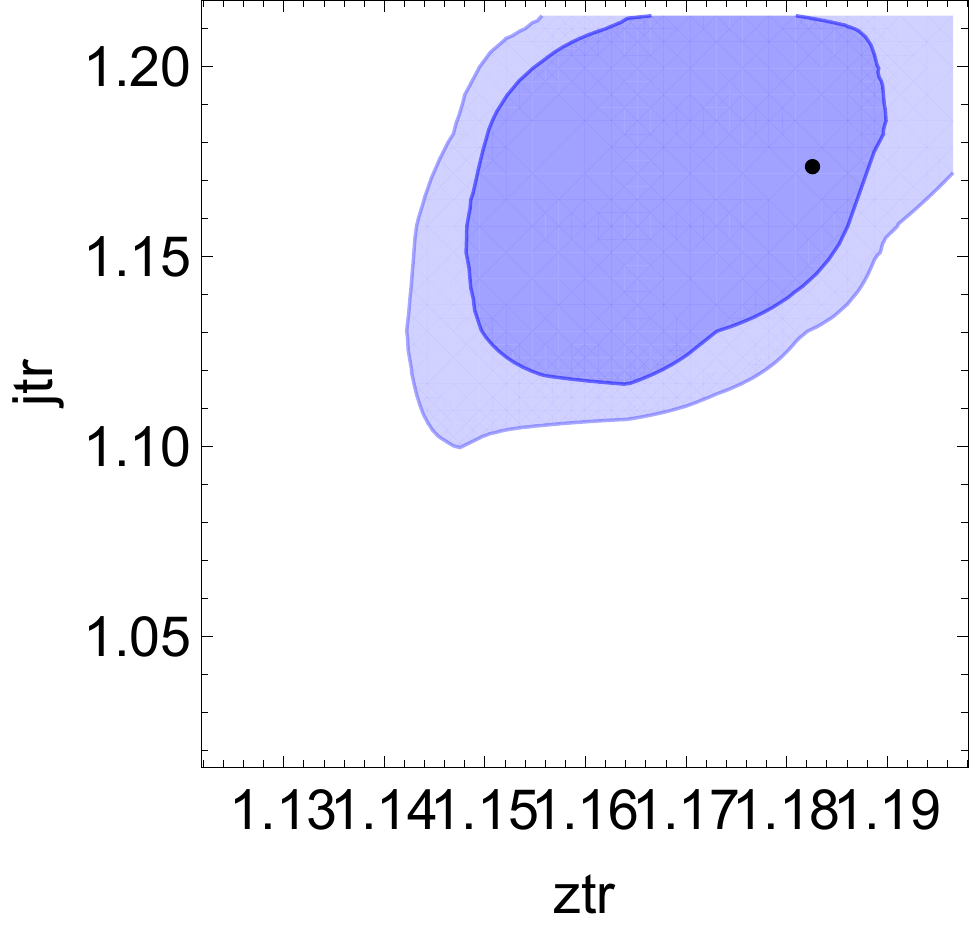}
\includegraphics[height=0.25\hsize,clip]{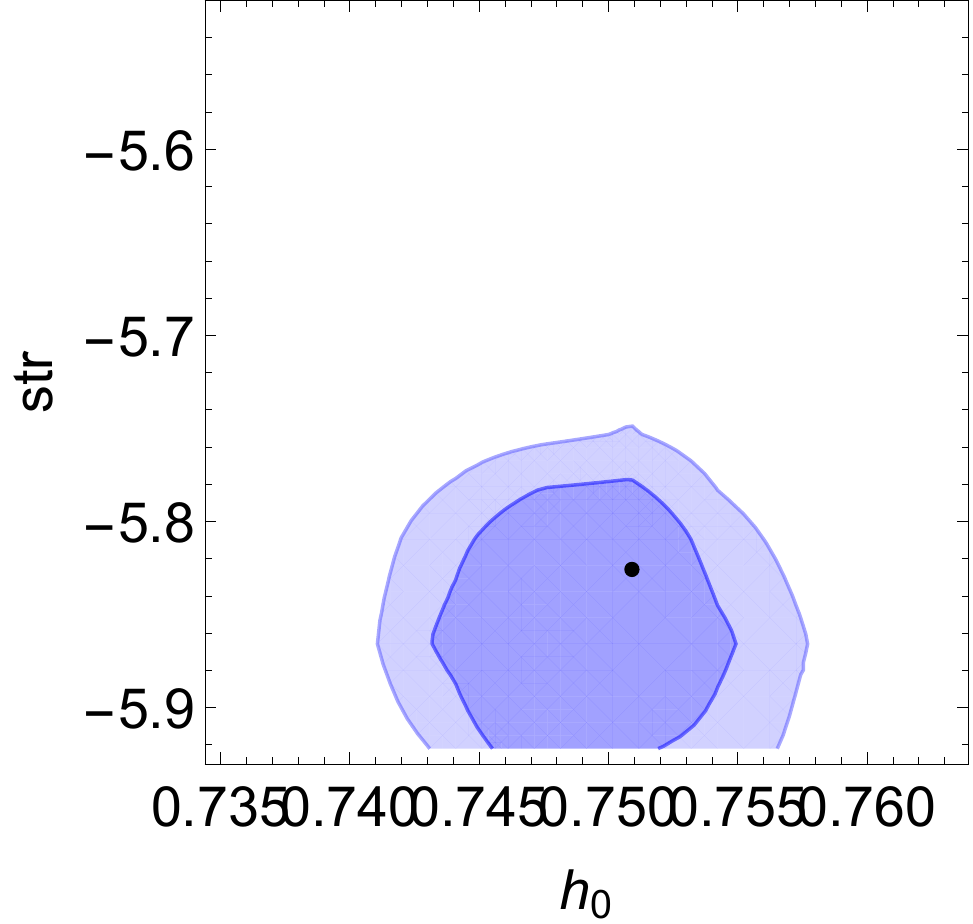}
\includegraphics[height=0.25\hsize,clip]{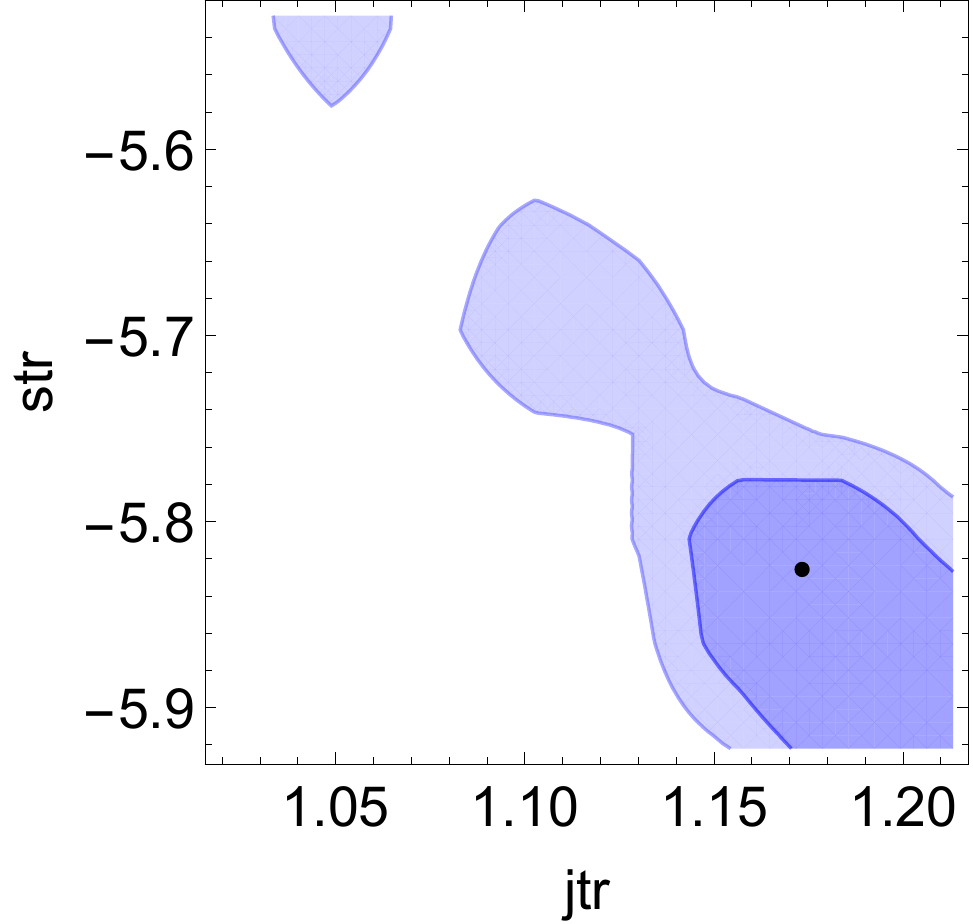}
\includegraphics[height=0.25\hsize,clip]{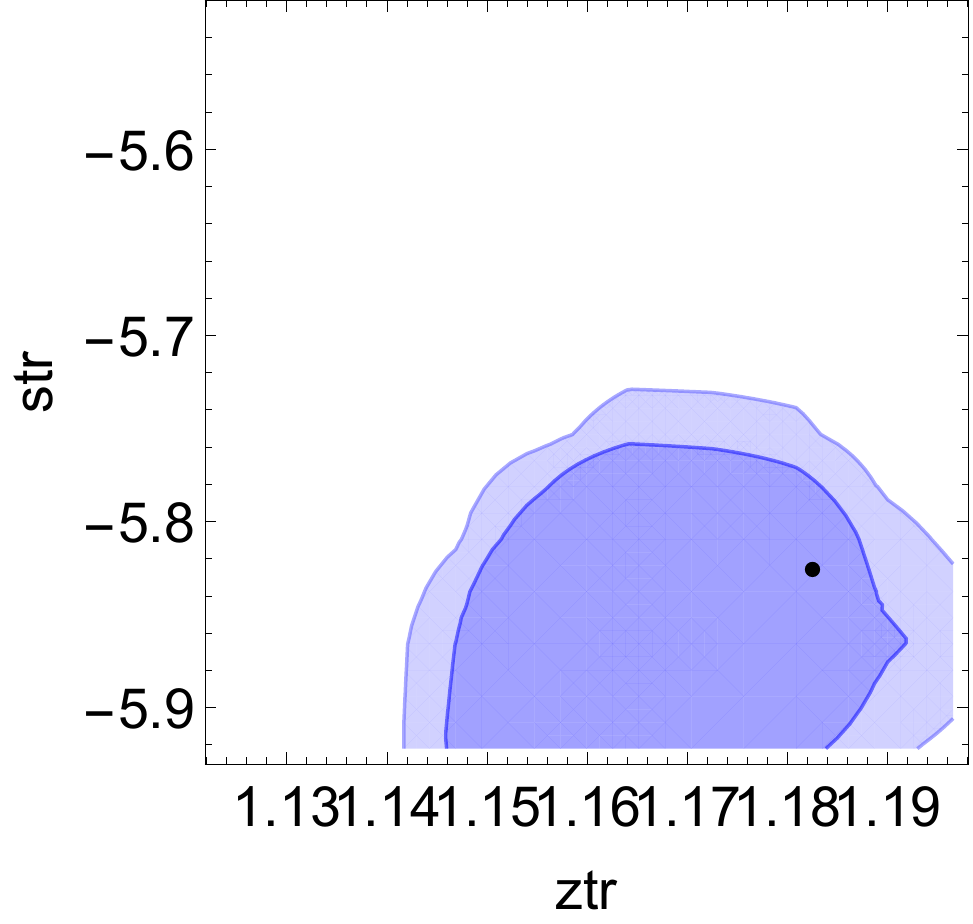}
\includegraphics[height=0.25\hsize,clip]{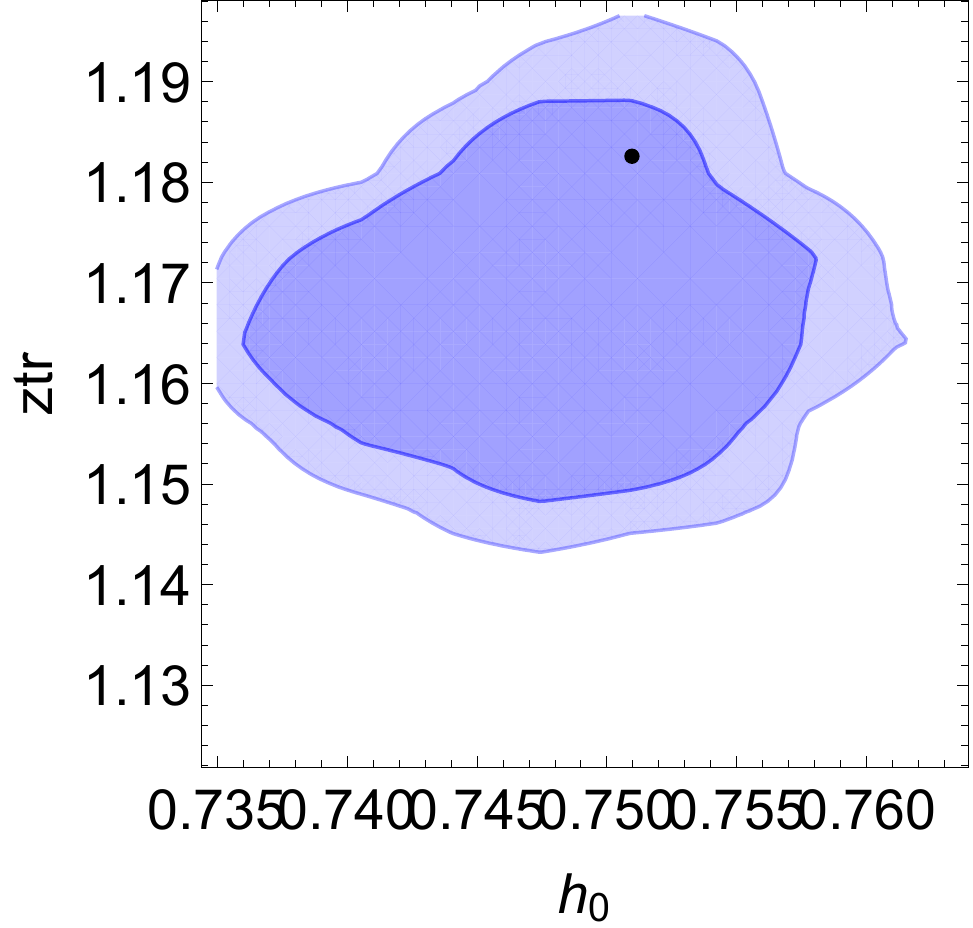}
\caption{Contours for the DDPE method using the SN+OHD+BAO data sets and the second order expansion of the deceleration parameter. }
\label{contourDDPEq2SNOHDBAO}
\end{figure*}

\noindent The contours evaluated using SNe Ia and OHD with BAO data sets are reported in Figs. \ref{contourDDPEq1SNOHDBAO} and \ref{contourDDPEq2SNOHDBAO}.

\end{document}